%%%%%%%%%%%%%%%%%%%%%%%%%%%%%%%%%%%%%%%%%%%%%%%%%%%%%%%%%%%%%%%%%%
%                                                                %
%  IMPORTANT:   If you do not have the figures or do             %
%               not have epsf.tex then uncomment                 %
%               the line indicated below.                        %
%                                                                %
%%%%%%%%%%%%%%%%%%%%%%%%%%%%%%%%%%%%%%%%%%%%%%%%%%%%%%%%%%%%%%%%%%

\newif\iffigs\figstrue
%  Uncomment the next line if you don't want the figures:
%\figsfalse

\input harvmac.tex
\iffigs
  \input epsf
\else
  \message{No figures will be included. See TeX file for more
information.}
\fi

\noblackbox

\batchmode
  \font\bbbfont=msbm10
\errorstopmode
\newif\ifamsf\amsftrue
\ifx\bbbfont\nullfont
  \amsffalse
\fi
\ifamsf
\def\IR{\hbox{\bbbfont R}}
\def\IZ{\hbox{\bbbfont Z}}
\def\IF{\hbox{\bbbfont F}}
\def\IP{\hbox{\bbbfont P}}
\else
\def\IR{\relax{\rm I\kern-.18em R}}
\def\IZ{\relax\ifmmode\hbox{Z\kern-.4em Z}\else{Z\kern-.4em Z}\fi}
\def\IF{\relax{\rm I\kern-.18em F}}
\def\IP{\relax{\rm I\kern-.18em P}}
\fi

\def\np#1#2#3{Nucl. Phys. B {#1} (#2) #3}
\def\pl#1#2#3{Phys. Lett. B {#1} (#2) #3}

\def\prl#1#2#3{Phys. Rev. Lett. {#1} (#2) #3}
\def\physrev#1#2#3{Phys. Rev. D {#1} (#2) #3}

\def\cF{{\cal F}}
\def\cA{{\cal A}}
\def\cK{{\cal K}}
\def\sign{{\rm sign}}
\def\CW{{\cal W}}
\def\c{c_{cl}}
\def\Tr{\mathop{\rm Tr}}
\def\lfm#1{\medskip\noindent\item{#1}}
\lref\niesem{A.J. Niemi and G.W. Semenoff, ``Axial-Anomaly-Induced
Fermion Fractionization and Effective Gauge-Theory Actions in
Odd-Dimensional Space-Times,'' \prl{51}{1983}{2077}.}
\lref\fdfp{N. Seiberg, ``Five Dimensional SUSY Field Theories,
Non-Trivial Fixed Points, and String Dynamics,'' hep-th/9608111,
\pl{388}{1996}{753}.}
\lref\sdfp{N. Seiberg, ``Non-trivial Fixed Points of The
Renormalization Group in Six Dimensions,'' hep-th/9609161,
\pl{390}{1996}{169}.}
\lref\msdelp{D.R. Morrison and N. Seiberg, ``Extremal Transitions and
Five-Dimensional Supersymmetric Field Theories,'' hep-th/9609070,
\np{483}{1996}{229}.}
\lref\dkvdelp{M.R. Douglas, S. Katz, and C. Vafa, ``Small Instantons,
del Pezzo Surfaces and Type I$'$ theory,'' hep-th/9609071.}
\lref\ewsmall{E. Witten, ``Small Instantons in String Theory,''
hep-th/9511030, \np{460}{1996}{541}.}
\lref\bds{T. Banks, M.R. Douglas, and N. Seiberg, ``Probing $F$-Theory
With Branes,'' hep-th/9605199, \pl{387}{1996}{278}.}
\lref\aps{P.C. Argyres, M.R. Plesser, and N. Seiberg,
``The Moduli Space of Vacua of $N=2$ SUSY QCD and Duality in $N=1$ SUSY
QCD,'' hep-th/9603042, \np{471}{1996}{159}.}
\lref\swsd{N. Seiberg and E. Witten, ``Comments on String
Dynamics in Six Dimensions,'' hep-th/9603003, \np{471}{1996}{121}.}
\lref\gh{O.J. Ganor and A. Hanany, ``Small $E_8$ Instantons and
Tensionless Noncritical Strings,'' hep-th/9602120, \np{474}{1996}{122}.}
\lref\gms{O.J. Ganor, D.R. Morrison, and N. Seiberg, ``Branes,
Calabi--Yau Spaces, and Toroidal Compactification of the $N{=}1$
Six-Dimensional $E_8$ Theory,'' hep-th/9610251, \np{487}{1997}{93}.}
\lref\redlich{A.N. Redlich, ``Gauge Noninvariance and Parity Violation of
Three-Dimensional Fermions,'' \prl{52}{1984}{18};
``Parity Violation and Gauge Non-invariance of the Effective Gauge Field
Action in Three Dimensions,'' \physrev{29}{1984}{2366}.}
\lref\dlp{M.J. Duff, H. Lu, and C.N. Pope, ``Heterotic Phase
Transitions and Singularities of the Gauge Dyonic String,''
hep-th/9603037, \pl{378}{1996}{101}.}
\lref\wittanom{E. Witten, ``An $SU(2)$ Anomaly,'' \pl{117}{1982}{324}.}
\lref\dewlvp{B. de Wit, P. Lauwers, and A. Van Proeyen, ``Lagrangians
of $N=2$ Supergravity-Matter Systems,'' \np{255}{1985}{269}.}
\lref\agw{L. Alvarez-Gaum\'e and E. Witten, ``Gravitational Anomalies,''
\np{234}{1983}{269}.}
\lref\mandf{E. Witten, ``Phase Transitions in $M$-Theory and
$F$-Theory,'' hep-th/9603150, \np{471}{1996}{195}.}
\lref\mdgm{M.R. Douglas and G. Moore, ``D-branes, Quivers, and ALE
Instantons,'' hep-th/9603167.}
\lref\dgdb{M.R.~Douglas, ``Gauge Fields and D-branes,''
hep-th/9604198.}
\lref\jhgm{J.A. Harvey and G. Moore, ``Algebras, BPS States and Strings,''
hep-th/9510182, \np{463}{1996}{315}.}
\lref\BSV{M. Bershadsky, V. Sadov, and C. Vafa, ``D-Strings on
D-Manifolds,'' hep-th/9510225, \np{463}{1996}{398}.}
\lref\aspinwall{P.S. Aspinwall,
``Enhanced Gauge Symmetries and Calabi--Yau
Threefolds,'' hep-th/9511171, \pl{371}{1996}{231}.}
\lref\KM{A. Klemm and P. Mayr, ``Strong Coupling Singularities and
Non-Abelian Gauge Symmetries in $N=2$ String Theory,'' hep-th/9601014,
\np{469}{1996}{37}.}
\lref\KMP{S. Katz, D.R. Morrison, and M.R. Plesser, ``Enhanced Gauge
Symmetry in Type II String Theory,'' hep-th/9601108, \np{477}{1996}{105}.}
\lref\MVi{D.R. Morrison and C. Vafa, ``Compactifications of F-Theory on
Calabi--Yau Threefolds, I,'' hep-th/9602114, \np{473}{1996}{74}.}
\lref\MVii{D.R. Morrison and C. Vafa, ``Compactifications of F-Theory on
Calabi--Yau Threefolds, II,'' hep-th/9603161, \np{476}{1996}{437}.}
\lref\AG{P.S. Aspinwall and M. Gross, ``The $SO(32)$ Heterotic String on a
K3 Surface,'' hep-th/9605131, \pl{387}{1996}{735}.}
\lref\BKKM{P. Berglund, S. Katz, A. Klemm, and P. Mayr, ``New Higgs
Transitions between Dual $N{=}2$ String Models,'' hep-th/9605154,
\np{483}{1996}{209}.}
\lref\BIKMSV{M. Bershadsky, K. Intriligator, S. Kachru, D.R. Morrison,
V. Sadov, and C. Vafa, ``Geometric Singularities and Enhanced Gauge
Symmetries,'' hep-th/9605200, \np{481}{1996}{215}.}
\lref\sadov{V. Sadov, ``Generalized Green--Schwarz
Mechanism in F-Theory,''
hep-th/9606008, \pl{388}{1996}{45}.}
\lref\KV{S. Katz and C. Vafa, ``Matter From Geometry,'' hep-th/9606086.}
\lref\rKMnew{S. Katz and D.R. Morrison, to appear.}
\lref\ccdf{A.C. Cadavid, A. Ceresole, R. D'Auria, and S. Ferrara,
``Eleven-Dimensional Supergravity Compactified on a Calabi--Yau
Threefold,'' hep-th/9506144, \pl{357}{1995}{76}.}
\lref\fkm{S. Ferrara,
R.R. Khuri, and R. Minasian, ``M-Theory on a Calabi--Yau Manifold,''
hep-th/9602102, \pl{375}{1996}{81}.}
\lref\fms{S. Ferrara,
R. Minasian, and A. Sagnotti, ``Low Energy Analysis of $M$ and $F$
Theories on Calabi--Yau Threefolds,'' hep-th/9604097,
\np{474}{1996}{323%--342}.}
}.}
\lref\kleiman{S. Kleiman, ``Toward a Numerical Theory of Ampleness,''
Annals of Math. {\bf 84} (1966) 293.}
\lref\kawamata{Y. Kawamata, ``Crepant Blowing-Up of $3$-Dimensional
Canonical Singularities and its
Application to Degenerations of Surfaces,''
Annals of Math. {\bf 127} (1988) 93.}
\lref\rGMnew{A. Grassi and D.R. Morrison, unpublished.}
\lref\rBirGeoRDP{D.R. Morrison, ``The Birational Geometry of Surfaces with
Rational Double Points,'' Math. Ann. {\bf 271} (1985) 415%--438.}
.}
\lref\grauert{H. Grauert, ``\"Uber Modifikationen und exzeptionelle
analytische Mengen,'' Math. Ann. {\bf 146} (1962) 331%--368.}
.}
\lref\mumford{D. Mumford, ``The Topology of Normal Singularities of an
Algebraic Surface and a Criterion for Simplicity,''
Inst. Hautes \'Etudes Sci. Publ. Math. {\bf 9} (1961) 5%--22.}
.}
\lref\AGM{P.S. Aspinwall, B.R. Greene, and D.R. Morrison, ``Calabi--Yau
Moduli Space, Mirror Manifolds and Spacetime Topology Change in String
Theory,'' hep-th/9309097, \np{416}{1994}{414}.}
\lref\flops{J. Koll\'ar, ``Flops,'' Nagoya Math. J. {\bf 113} (1989) 15.}
\lref\reid{M. Reid, ``Canonical Threefolds,'' Journ\'ees de G\'eometrie
Alg\'ebrique d'Angers, Sijthoff \& Nordhoff, 1980, p. 273.}
\lref\mori{S. Mori, ``Threefolds Whose Canonical Bundles Are Not
Numerically Effective,'' Annals of Math. {\bf 116} (1982) 133.}
\lref\kawcone{Y. Kawamata, ``The Cone of Curves of Algebraic Varieties,''
Annals of Math. {\bf 119} (1984) 603.}

\Title{hep-th/9702198, RU-96-99, IASSNS-HEP-96/112}
{\vbox{\centerline{Five-Dimensional Supersymmetric Gauge Theories}
\vskip2pt\centerline{and Degenerations of Calabi--Yau Spaces}}}
\bigskip
\centerline{Kenneth Intriligator$^{1}$\footnote{${}^*$}{On leave
1996--1997
{}from Department of Physics, University of California, San Diego.},
David R. Morrison$^{2,1}$\footnote{${}^\dagger$}{On leave 1996--1997
{}from Department of Mathematics, Duke University.},
and Nathan Seiberg$^{3}$}
\vglue .5cm
\centerline{$^1$School of Natural Sciences, Institute for Advanced
Study, Princeton, NJ 08540, USA}
\centerline{$^2$School of Mathematics, Institute for Advanced Study,
Princeton, NJ 08540, USA}
\centerline{$^3$Department of Physics and Astronomy, Rutgers University,
Piscataway, NJ 08855, USA}
\vglue.8cm

\noindent

We discuss five-dimensional supersymmetric gauge theories.  An anomaly
renders some theories inconsistent and others consistent only upon
including a Wess--Zumino type Chern--Simons term.  We discuss some
necessary conditions for existence of nontrivial renormalization group
fixed points and find all possible gauge groups and matter content which
satisfy them.  In some cases, the existence of these fixed
points can be inferred {}from string duality considerations.  In other
cases,
they
arise {}from M-theory on Calabi--Yau threefolds.  We explore connections
between aspects of the gauge theory and Calabi--Yau geometry.
A consequence of our classification of field theories with nontrivial
fixed points is a fairly complete classification of a class of
singularities of Calabi--Yau threefolds which generalize the
``del Pezzo contractions'' and occur at higher
codimension walls of the K\"ahler cone.

\Date{2/97}

\newsec{Introduction}

It has recently been argued that string theory duality predicts the
existence of new, nontrivial fixed points of the renormalization group
for a variety of field theories in a variety of dimensions.  The field
theories arise on D-brane probes in string theory, whose dynamics must
reflect aspects of string duality \refs{\mdgm, \dgdb, \bds}.  For
example, in \fdfp\ it was thus argued that there are nontrivial fixed
points in five-dimensional supersymmetric gauge theories.  In
particular, it was argued in \fdfp\ that five-dimensional
supersymmetric $SU(2)$ gauge theory has strongly coupled nontrivial
fixed points for $n_F\leq 7$ quarks in the fundamental representation.
In \refs{\fdfp,
\sdfp} it was conjectured, more generally, that nontrivial fixed points
exist when a certain one-loop quantity has a particular sign, a
generalization to higher dimensions of the notion of asymptotic freedom.
We will discuss this quantity and the conditions for a nontrivial fixed
point for general five-dimensional supersymmetric gauge theories.  We
provide a rather complete classification of all possible nontrivial
five-dimensional fixed points of gauge-theoretic origin.

As discussed in \refs{\msdelp ,\dkvdelp}, there is a direct
correspondence between the nontrivial fixed points which exist for
five-dimensional $SU(2)$ gauge theory with $n_F\leq 7$ quarks and the
mathematical classification of a particular kind of singularity which
can occur in Calabi--Yau threefolds at codimension-one boundary walls
of the K\"ahler cone: the collapse of a ``del Pezzo surface'' with
$n_P\leq 7$ blown-up points.  The correspondence follows via
compactification of M-theory to five dimensions.  The generalization
to higher codimension boundary walls involves the study of the class
of singularities known as {\it isolated canonical
singularities}. Although quite a bit is known about the structure of
these singularities \reid, to give a complete classification would
appear to be a very challenging mathematical problem.  However, for
those isolated canonical singularities which arise {}from contraction
of a curve of singularities (and so correspond to the strong coupling
limit of a gauge theory) we can and do give a fairly complete
classification.  Our classification omits cases such as the $E_0$
theory of \msdelp\ which do not arise in this way.

In F-theory on a Calabi--Yau threefold to six dimensions there are
strings which couple to two-form gauge fields $B_{\mu \nu}$ and can
become tensionless at singularities on the moduli space of the
Calabi--Yau \refs{\swsd, \dlp}.  For example, this is the situation
with the small $E_8$ instanton
\refs{\gh, \swsd}. Upon reduction to five dimensions of the small $E_8$
instanton, the two-form $B_{\mu \nu}$ yields a gauge field $A_\mu$,
which is part of an $SU(2)$ gauge group which has $n_F=7$ flavors.  In
other cases, there can be, already in six dimensions, non-Abelian
gauge fields in addition to the two-form $B_{\mu  \nu}$.  It is an
interesting possibility that these gauge fields could combine in five
dimensions with that obtained upon reduction of the $B_{\mu \nu}$
into a larger non-Abelian gauge group, leading to fixed points of the
type considered in this paper.

In the next section, we discuss general aspects of five-dimensional
supersymmetric gauge theories and argue that there is an anomaly which
renders some theories inconsistent and other theories consistent only
upon including a Chern--Simons term.  Some of the results about the
anomaly have already appeared in \niesem.  In sect.~2.3 we present a
general expression, valid for arbitrary gauge group and matter
content, for the exact effective prepotential on the Coulomb branch.

In sect.~3 we discuss the general necessary condition for there to be
a nontrivial renormalization group fixed point: the prepotential
should be a convex function over the entire Coulomb branch.  This is
the generalization to higher-rank gauge groups of the condition
discussed in \fdfp\ for the $SU(2)$ case.

In the following few sections we classify the various gauge groups and
matter content which satisfy this condition.

In sect.~4 we consider $Sp(N)$ gauge theory.  Because the fundamental
representation is pseudo-real, classically half-hypermultiplets in the
fundamental are possible.  However, the anomaly implies that the
theory is only consistent when the number of hypermultiplets is
integral.  We argue that there can be nontrivial fixed points only
with matter in the antisymmetric tensor and/or fundamental
representations, with $n_A\leq 1$ antisymmetric tensors; for $n_A=1$,
the number of fundamentals is $n_F\leq 7$, while for $n_A=0$ it is
$n_F\leq 2N+4$.  The existence of nontrivial fixed points for $n_A=1$
or $n_A=0$ with $n_F\leq 7$ follows {}from stringy probe
considerations, as in \fdfp.  The existence of the other cases is
demonstrated in sect.~9.

In sect.~5 we consider $SU(N)$ gauge theory.  Here we can, and
sometimes {\it must}, add a bare cubic prepotential with coefficient
$c_{cl}$.  The anomaly restricts $c_{cl}$; for example, with an odd
number of matter flavors in the fundamental representation, $c_{cl}$
must be half of an odd integer, and thus can not be zero.  We classify
the theories which satisfy the necessary convexity condition for a
non-trivial fixed point.  For general $N$, there can only be matter in
the fundamental representation, with the number of flavors and
$c_{cl}$ restricted by $n_F+2|\c |\leq 2N$.  For $N\leq 8$ there can
also be fixed points with $n_A=1$ matter field in the antisymmetric
tensor representation and $n_F$ in the fundamental provided $n_F+2|\c
|\leq 8-N$.  For $N=4$ there can also be a fixed point with $n_A=2$
matter fields in the antisymmetric tensor representation and $n_F=\c
=0$.

In sect.~6 we consider $Spin(M)$ gauge theories, showing that there can
be fixed points with $n_V\leq M-4$ matter fields in the vector
representation.  For $M\leq 12$ there can also be fixed points with
$n_S\leq 2^{6-\half M}$ ($n_S\leq 2^{5-\half (M-1)}$) matter fields in
the spinor representation for $M$ even (odd) and $n_V\leq M-4$.

In sect.~7 we consider the exceptional gauge groups.  For $G_2$ there
can be a nontrivial fixed point for $n_7\leq 4$ fundamentals.  For
$F_4$, there can be a fixed point for $n_{26}\leq 3$ fundamentals.
For $E_6$, there can be a fixed point for $n_{27}\leq 4$ fundamentals.
For $E_7$, there can be a fixed point for $n_{56}\leq 3$ ($ n_{56}$
can be half-integral) fundamentals.  For $E_8$, there can only be a
fixed point for the theory without matter fields.

In sect.~8 we discuss the Calabi--Yau interpretation of these results.
The connection is made via $M$ theory.  Determining a detailed
correspondence between configurations of surfaces on Calabi--Yau
threefolds and gauge theories with specified matter content occupies
us for the bulk of this section.  We compute the prepotential {}from
the Calabi--Yau side, and check that it agrees with the one calculated
in gauge theory.  This leads to some miraculous formulas for the cubic
intersection form among certain divisors on Calabi--Yau threefolds, as
anticipated in \jhgm.

Finally, in sect.~9 we study strong coupling limits of the Calabi--Yau
theories.  We are not able to settle the question of the existence of
a strong coupling limit in all cases, but we do find examples in which
such limits exist, enabling us to conclude that many of the exotic
strong-coupling points---such as $Sp(N)$ with $n_A=0$ and
$n_F\leq 2N+4$---do in fact occur.

\newsec{General Aspects of Supersymmetric Gauge Theories in Five
Dimensions}

On the Coulomb branch of the moduli space, the real scalar $\Phi$ of
the vector multiplet $\cA$ gets expectation values in the Cartan
sub-algebra of the gauge group $G$, breaking $G$ to the Cartan
$U(1)^r$, with $r=$rank$(G)$.  The Coulomb branch is thus a Weyl
chamber, which is a {\it wedge}\/ subspace of $\IR ^r$ parameterized
by Cartan scalars $\phi^i$ in $\IR ^r/\CW$ ($\CW$ is the Weyl group of
$G$), the expectation values of the massless Cartan $U(1)^r$ vector
multiplets $\cA ^i$.  Away {}from the origin on the Coulomb branch the
low-energy effective Abelian theory is characterized by a prepotential
$\cF (\cA ^i)$, which is at most cubic locally in the $\cA ^i$.  There
is enhanced gauge symmetry on the walls of the Weyl chamber.

There can also be matter hypermultiplets, ``quarks,'' in a
representation $\oplus _f {\bf r}_f$, where ${\bf r}_f$ are
irreducible representations of $G$.  In addition to the Coulomb
branch, there can be a Higgs branch associated with expectation values
of the scalars in the matter hypermultiplets.  The Higgs branch is
hyper-K\"ahler and generally intersects the Coulomb branch along a
subspace where certain $\phi ^i$, those associated with photons which
get Higgsed, are set to zero.  And, as in four dimensions \dewlvp,
there is a lack of neutral couplings between vector multiplets and
hypermultiplets which implies that the gauge coupling eigenvalues for
the remaining $\phi ^i$ are independent of the expectation value on
the Higgs branch.  Also, as in four dimensions \aps, because the gauge
coupling can be regarded as a background expectation value of a vector
multiplet, the Higgs branch does not receive quantum corrections.  We
thus focus on the Coulomb branch.

An important note about kinematics in five dimensions is in order
here.  One symmetry of our problem is parity ${\cal P}: x^\mu
\rightarrow -x^\mu$, combined with
changing the sign of the scalars (and mass terms).  When these
theories are considered as dimensional reductions of the
six-dimensional theory, this symmetry is understood as part of the
six-dimensional Lorentz group.  Also, there is a charge conjugation
symmetry, $\cal{C}$.  Mass terms, which are like background gauge
fields, change sign under charge conjugation.  Thus, with non-zero
masses, only the product $\cal{CP}$ is a symmetry of the classical
Lagrangian.

Finally, we note that for the particular case of gauge group $G=Sp(N)$
with no massless matter fields in the fundamental representation,
there is an additional discrete parameter needed to specify the
quantum theory.  As in \dkvdelp, this parameter can be interpreted as
a $Z_2$ valued theta angle associated with $\pi _4(Sp(N))=Z_2$; this
is analogous to the theta angle of 4d gauge theories, which is
associated with $\pi _3(G)$.  When there are massless matter fields in
the fundamental representation of $Sp(N)$, the discrete parameter is
not physical, as the theta angle can then be rotated away.  When the
matter fields in the fundamental representation are massive we can
replace this parameter with the sign of the determinant of the mass
matrix. For gauge
groups other than $Sp(N)$, $\pi _4$ is trivial and thus there is no
theta parameter.

\subsec{Abelian gauge theory and the Wess--Zumino term}

The prepotential of a $U(1)^r
$ gauge theory in
five dimensions is a cubic polynomial in $\phi^i$.  In every open set
in the moduli space it is of the form
\eqn\uonepre{\CF = {t^{(0)}_{ij} \over 2} \phi^i \phi^j + {c_{ijk}
\over 6} \phi^i \phi^j \phi^k.}
In different open sets the coefficients $t^{(0)}$ and $c$ can be
different but
\eqn\adtdef{\eqalign{
&{a_D}_i=\partial_i \CF \cr
&t(\phi)_{ij}=\partial_i \partial_j \CF \cr}}
must be continuous.  $\cF$ leads to an effective
gauge coupling proportional to
\eqn\uonet{t(\phi)_{ij}F^i F^j,}
a metric on the moduli space proportional to
\eqn\uonemetric{(ds)^2=t(\phi)_{ij}d\phi^i d\phi^j,}
and a Chern--Simons term
\eqn\uonecs{{c_{ijk}\over 24\pi^2} A^i \wedge F^j \wedge F^k.}
Our definition of $c_{ijk}$ here differs by a factor of two {}from the
conventions in \refs{\fdfp, \msdelp,\gms}.

Gauge invariance restricts the coefficients $c_{ijk}$ because \uonecs\
is not gauge invariant but the integral of \uonecs\ should be well
defined modulo $2\pi i \ell$, with $\ell\in \IZ$.  For the special case of
$r=1$ (one-dimensional moduli space) gauge invariance on a generic
five-manifold thus quantizes $c \in 6 \IZ$.  However, on the special
manifolds which occur in M-theory \mandf
\eqn\uonecqu{c \in \IZ.}
For any $r$, the condition of $c_{ijk}$ can be obtained by writing the
integral of \uonecs\ over a five-manifold $W$ as the integral over a
six-manifold $Y$ with $W=\partial Y$,
\eqn\intoversm{2\pi {c_{ijk}\over 6}\int _{Z_6}({F^i\over 2\pi})
\wedge ({F^j\over 2\pi})\wedge ({F^k\over 2\pi})\equiv
2\pi i{c_{ijk}\over 6}c_1(L^i)c_1(L^j)c_1(L^k),} which should be $2\pi
i\ell$, with $\ell\in \IZ$.  The quantization condition \uonecqu\ follows
because $c_1(L)^3$ is divisible by six for the special manifolds which
occur in M-theory \mandf.  More generally, for $U(1)^r$, $c_1(L)^3$ is
divisible by six for any $L$; writing $L=L_i+L_j+L_k$ and expanding,
it follows that $c_1(L_i)^3$ is divisible by six, $c_1(L_i)^2c_1(L_j)$
is divisible by two for $i\neq j$, and $c_1(L_i) c_1(L_j)c_1(L_k)$ is
integral for $i\neq j\neq k$. Taking combinatoric factors into
account, this gives for the quantization condition
\eqn\abcijk{c_{ijk}\in \IZ\qquad\hbox{for any}\ i,j,k.}

For a $U(1)^r$ gauge theory with no matter, quantum effects do not
change the prepotential, so $\cF =\cF _{classical}$.

Consider now a gauge theory coupled to matter fields and study a
neighborhood in the Coulomb branch of its moduli space including the
point where the matter fields are massless.  At a generic point in
this neighborhood the prepotential is as in \uonepre.  The cubic terms
in the prepotential are a sum of a term which exists at tree level and
a quantum term which is generated by loops.  The quantum contribution
is not smooth in this neighborhood---it has singularities at points
where new massless particles exist.  However, the classical part must
be smooth.  Their sum must respect the quantization conditions
discussed above.

For example, consider a $U(1)$ gauge theory with $n_f$ ``electrons''
which are massless at $\phi=0$.  The quantum contribution to the
prepotential was computed in \mandf
\eqn\uonenfq{c_{quantum}=-{n_f \over 2} \sign(\phi).}
Combining this with a classical cubic term $c_{classical}$ we learn
{}from \uonecqu\ that for consistency
\eqn\consis{c_{classical} -{n_f \over 2} \in \IZ}
and in particular, for odd $n_f$ $c_{classical} $ cannot vanish.
Thus, for odd $n_f$, the theory necessarily has $\cal{C}$ and
$\cal{P}$ broken, with $\cal{CP}$ preserved.  It is easy to generalize
to the case of electrons with masses $m_f$
\eqn\uonenfqm{c_{quantum}=-\half \sum_{f=1}^{n_f}  \sign(\phi + m_f).}
Consider integrating out $n$ massive electrons with mass $m >0$.  They
induce $c_{induced}= -{n\over 2}$, which appears as a classical term in
the low-energy theory of the $n_f-n$ massless electrons.  Hence, one
can interpret $c_{classical}$ as being induced by loops of massive
electrons.

Another way of expressing this phenomenon is the following.  The
fermion determinant in a $U(1)$ gauge theory with odd number of
flavors is not gauge invariant.  It is multiplied by $-1$ under
certain gauge transformations.  The Chern--Simons term with half-integer
$c_{classical}$ plays the role of a Wess--Zumino term.  Its
lack of gauge invariance compensates for the lack of gauge invariance
of the fermion determinant.

This understanding is in accord with a phenomenon noticed in \gms\ when
the five-dimensional theory was reduced on a circle to four dimensions.
The moduli space of the four-dimensional theory is a cylinder.  The
monodromy around the point with $n_f$ massless electrons is $T^{n_f}$.
The monodromies around the two circles of the cylinder are $T^{(n_f/ 2)
+c}$ and $T^{(n_f/ 2) -c }$.  For consistency, $(n_f/ 2) +c \in \IZ$ \gms.
This provides an independent derivation of our anomaly.

Note that in \uonenfq, because $\phi$ is odd under both $\cal{P}$ and
$\cal{C}$, the induced Chern--Simons term properly respects $\cal{P}$
and $\cal{C}$ if the classical theory does, $c_{classical}=m=0$.
On the other hand, with $m\neq 0$ only $\cal{CP}$ is a symmetry, which
is exhibited in the low-energy theory with the massive matter
integrated out by the non-zero constant $c_{induced}$.  (Because there
can be a $c_{classical}$ which is tuned to cancel $c_{induced}$, two
classical violations of $\cal{C}$ and $\cal{P}$ can cancel in the
infrared.)

For $G=U(1)^r $, and matter fields of charges $(q_f)_i$ under the
$U(1)_i$, the above have obvious generalizations; for example, the
generalization of \uonenfqm\ is
\eqn\uonesnfq{(c_{quantum})_{ijk}=-\half \sum_{f=1}^{n_f}(q_f)_i(q_f)_j
(q_f)_k\sign((q_f)_r\phi ^r+m_f).}

\subsec{Non-Abelian gauge theory and the Wess--Zumino term}

We now consider a non-Abelian simple gauge group $G$ with matter,
``quarks'' in a representation $\oplus_f {\bf r} _f$ where ${\bf r}
_f$ are irreducible representations of $G$ with masses $m_f$. The
classical prepotential is
\eqn\nonabprec{{m_0\over 2} \Tr \Phi^2 + {c_{classical} \over 6} \Tr
\Phi^3,}
where $m_0=1/g_{cl}^2$ is the classical gauge coupling; it has
dimensions of mass in a five-dimensional theory.  Take $\Phi= \phi^a
T_a$, with the index $a$ running over the generators of $G$ which are
represented by the matrices $T_a$ in the fundamental representation of
$G$.  The cubic term in \nonabprec\ leads to terms in the Lagrangian
proportional to
\eqn\nonabcou{c_{classical} d_{abc}\phi^a \partial \phi^b \partial
\phi^c+ c_{classical} d_{abc}\phi^a F^b F^c,}
with the $d$ symbol defined by
\eqn\defd{d_{abc}=\half \Tr T_a (T_bT_c + T_cT_b),}
and the non-Abelian Chern--Simons term
\eqn\nonabecs{{c_{classical}  \over 24\pi^2}\Tr (A\wedge F \wedge F -
{1 \over 2} A\wedge A \wedge A \wedge F + {1 \over 10} A \wedge A
\wedge A \wedge A \wedge A ).}
Such a $c_{classical}$ term is possible only for $G=SU(N)$, $N\geq 3$,
as those are the only (simple) gauge groups with a nontrivial third-order
Casimir.

As in the $U(1)$ example above, the cubic term in \nonabprec\ can be
interpreted as arising {}from integrating out massive quarks.  More
explicitly, for $G=SU(N)$ (with $N\geq 3$) integrating out $n_f$
fundamental matter fields of mass $m$ induces
\eqn\inducedc{c_{induced}=-sign(m){n_f\over 2}.}
(The generalization for integrating out massive matter in other
representations modifies \inducedc\ by a factor of the cubic index of
the representation.)  Consider now the $SU(N)$ theory with quarks, all
taken to be massive, in the fundamental representation.  If there are
$n_{f+}$ massive quark flavors with masses $m_f>0$ and $n_{f-}$ with
masses $m_f<0$, the low-energy effective $c$ is
\eqn\effec{c_{eff}=c_{classical}- {n_{f+}-n_{f-} \over 2}.}

The Chern--Simons term \nonabecs\ is not gauge invariant.  For gauge
transformations $g$ which are nontrivial in $\pi_5 (SU(N))= \IZ$, it
changes by
\eqn\varia{{c_{eff} \over 240\pi ^2} \int \Tr g^{-1}
dg g^{-1} dg g^{-1} dg
g^{-1} dgg^{-1} dg= c_{eff} 2 \pi i \ell,}
with $\ell \in \IZ$.  We see that for consistency we need
\eqn\conscsum{c_{eff}=c_{classical}- {n_{f+}-n_{f-} \over 2}\in \IZ.}
If we consider now taking the masses of the flavors to zero, the
numbers $n_{f\pm}$ become ambiguous but $n_{f+}-n_{f-}=n_f$ mod 2.
Therefore, in the theory with $n_f$ massless flavors
\eqn\conscsu{c_{eff}=c_{classical}- {n_f \over 2}\in \IZ.}
In particular, for odd number of flavors we must add the cubic term in
\nonabprec.  A similar phenomenon in three dimensions was 
observed in \refs{\redlich, \agw}; the generalization to five
dimensions had already been noted in \niesem. Note that unlike the
$U(1)$ case mentioned above, here we do not need the requirement of
special five-manifolds to get the correct quantization condition.

For gauge groups other than $SU(N)$, as there is no nontrivial
third-order Casimir, $c_{classical}=0$ and lack of gauge invariance
under a nontrivial element of $\pi _5(G)$ would render the theory
inconsistent.  The only other cases of nontrivial $\pi _5$ are
$Sp(N)$, which have $\pi _5(Sp(N))=\IZ _2$.  Correspondingly, there is
a global anomaly, similar to that of \wittanom, which implies that
$Sp(N)$ theories must have an even number of half-hypermultiplets in
the fundamental representation.  One way to see that is to consider a
$d=5$ instanton associated with $\pi _4(Sp(N))=Z_2$.  It has one
zero-mode for each half-hypermultiplet in the fundamental
representation; similar to the situation for the global anomaly in
four dimensions, this implies that the five-dimensional theory with an
odd number of half-hypermultiplets is inconsistent.  The requirement
that the number of half-hypermultiplets of $Sp(N)$ be even is also
consistent with compactification to four dimensions, where the number
of half-hypermultiplets has to be even because of $\pi _4(Sp(N))=\IZ
_2$.

\subsec{Low-energy effective theory}

Along the Coulomb branch the non-Abelian gauge group $G$ is broken to
$U(1)^r$, with $r=$rank$(G)$, and the theory is described by an
Abelian low-energy effective theory for the $\cA ^i$ in $\cA =\sum
_{i=1}^r\cA ^iT_i$, with $T_i$ the Cartan generators of $G$.  Five
dimensional gauge invariance implies that the exact quantum
prepotential $\cF (\cA ^i)$ is at most cubic in the $\cA_i$.
Indeed, only if $\partial _i\partial _j\partial _k \cF \equiv c_{ijk}$
is a constant, could the quantization condition
\abcijk\ possibly be satisfied.  Because it is at most cubic,
the exact low-energy effective prepotential is determined by a
one-loop calculation.  For arbitrary gauge group $G$ and matter
multiplets in representations ${\bf r}_f$, with masses $m_f$, the
result for the exact quantum prepotential (written for the scalar
components $\phi ^i$ of $\cA ^i$) is
\eqn\fgen{\cF=\half m_0h_{ij}\phi ^i\phi ^j +{c_{classical}\over 6}
d_{ijk}\phi ^i\phi ^j\phi ^k+ {1\over 12}\left(\sum _{{\bf R}}|{\bf R}
\cdot {\bf\phi} |^3-\sum _{f}\sum _{{\bf w}\in {\bf W}_f}|{\bf w}\cdot
{\bf \phi}+m_f|^3\right).}  Here $h_{ij}=\Tr (T_iT_j)$ and $d_{ijk}$
is \defd, both evaluated for the Cartan generators.  ${\bf R}$ are the
roots of $G$, and ${\bf W}_f$ are the weights of $G$ in the
representation ${\bf r}_f$.  The first two terms in \fgen\ are the
classical prepotential \nonabprec\ along the Coulomb branch.  The last
two terms in \fgen\ are the quantum contributions of the massive
charged vector and matter multiplets, respectively, which contribute
with opposite sign to the prepotential.  Explicit expressions for
\fgen\ will presented in later sections.

Note that \fgen\ agrees with what we find when we first integrate out
the massive matter to induce a cubic term in the non-Abelian theory
(it is only nontrivial for $G=SU(N)$ with $N\geq 3$) and then go along
the flat directions of the Coulomb branch.  In addition, there are
extra quadratic terms induced in the prepotential upon integrating out
massive flavors; these terms can always be canceled by a shift of
$m_0$ proportional to $m_f$.

Within the Weyl chamber for ${\bf \phi}$, each term ${\bf R}\cdot {\bf
\phi}$ in \fgen\
is either everywhere positive or everywhere negative, without changing
sign; the possible zeros are all at the boundaries of the Weyl
chamber.  The absolute value for these terms in \fgen\ thus simply
amount to assigning each term the appropriate sign throughout the Weyl
chamber.  On the other hand, it is possible for some terms ${\bf
w}\cdot {\bf
\phi} +m_f $ to change sign (even when $m_f=0$), within the Weyl chamber.
When that happens, because of the absolute value in \fgen, there are
different prepotentials in different sub-wedges of the Weyl chamber,
within which the terms ${\bf w}\cdot {\bf \phi} +m_f $ are of definite
sign; the boundaries of the sub-wedges are where some ${\bf w}\cdot
{\bf \phi} +m_f $ is zero.  At the boundaries of the sub-wedges, $\cF$
is not smooth but, because of the $|{\bf w}\cdot {\bf \phi} +m_f |^3$
dependence, $\cF$ and its Hessian, the metric $g^{-2}$, are continuous
across the boundaries of the sub-wedges.  Exactly as in \uonenfq, the
discontinuity in $\partial _i\partial _j\partial _k\cF$ is associated
with charged matter fields which become massless on the boundaries of
these sub-wedges.

In the explicit expressions for \fgen\ which will be given in later
sections, it can be seen that the low-energy effective Abelian theory
has a $(c_{eff})_{ijk}$ which satisfies the Abelian quantization
condition \abcijk\ in all cases as long as the non-Abelian theory does
not suffer {}from a global anomaly associated with $\pi _5(G)$.  Here it
is necessary to use a basis and normalization for the Cartan torus where
all Abelian charges, the roots and weights, are properly quantized to be
integers.  The contribution of the massive gauge fields, the sum over
the roots in \fgen, always leads to an integer contribution to $c_{ijk}$
for any gauge group.  For $G=SU(N)$ the appropriate basis for the Cartan
sub-algebra is generated by $(T^{(i)})^j_k=(\delta ^{i,j}-\delta
^{i+1,j}) \delta ^j_k$, $i=1, \dots N-1$, and the classical and matter
multiplet contribution to $c_{ijk}$ is
$c_{i,i,i+1}=-c_{i+1,i+1,i}=(c_{classical}-\half n_f)$.  The $c_{ijk}$
thus satisfy the Abelian quantization condition \abcijk\ precisely when
the non-Abelian theory satisfies the non-Abelian quantization condition
\conscsu.

Similarly, for $G=Sp(N)$ with $n_f$ hypermultiplets, i.e.\ $2n_f$
half-hypermultiplets, the contribution is $c_{iii}=-n_f$. The Abelian
quantization condition is satisfied since, as seen above, $n_f$ must be
an integer---the theories with an odd number of half-hypermultiplets
have a global anomaly.  For all other cases of gauge groups and matter
the effective $c_{ijk}$ satisfy \abcijk.

\newsec{Nontrivial Renormalization Group Fixed Points}

The metric $t(\phi)_{ij}=\partial _i\partial _j\cF$ should be
non-negative throughout the Weyl chamber of the Coulomb branch.  As
discussed in \fdfp, when the quantum contribution to the metric, which
is linear in $\phi$, is negative, at best the theory can be made
sensible in a subspace near the origin by taking the classical part
$t^{(0)}_{ij}=m_0h_{ij}>0$.  Eventually, away {}from the origin, the
negative quantum part becomes larger than $t^{(0)}_{ij}$, which is a
reflection of the fact that the theory is non-renormalizable and
eventually hits a Landau pole.  On the other hand, when the quantum
contribution to the metric is positive on the entire Coulomb branch
(Weyl chamber), it is possible to have a scale invariant fixed point
theory with $m_0=g_{cl}^{-2}=0$ \fdfp.  The theory is then sensible on
the entire moduli space of vacua.  A {\it necessary}\/ (though perhaps
not sufficient) condition \fdfp\ for a nontrivial strong coupling
fixed point is thus that the quantum part of the metric, $(\partial
_i\partial _j\cF )d\phi ^id\phi ^j$ must be non-negative throughout
the Weyl chamber.  We will examine general conditions on the gauge
theory for this to be the case.

The condition that the Hessian $\partial _i\partial _j \cF$ be a
positive matrix is equivalent to the condition that the prepotential
$\cF$ be a convex function throughout the Weyl chamber: for any two
points $\bf x$ and $\bf y$ in the Weyl chamber, the prepotential must
satisfy $\cF (\lambda {\bf x}+(1-\lambda ){\bf y})\leq \lambda \cF
({\bf x})+(1-\lambda )\cF ({\bf y})$ for $0\leq
\lambda\leq 1$.  (A function which satisfies the opposite inequality
is referred to as concave.)

We will investigate the conditions required on the gauge group and
matter content for the prepotential \fgen\ with $m_0=0$ to be a convex
function on the entire Weyl chamber.  Because $\cF$ and $\partial _i\cF$
vanish at the origin, a consequence of the convexity of $\cF$ is that
$\cF\geq 0$ throughout the Weyl chamber Coulomb branch.

It is easily seen that the term in
\fgen\ associated with the vector multiplets is purely convex:  each
term in the sum on ${\bf R}$ is obviously convex as a function of
${\bf R}\cdot \phi$, and the sum of convex functions is convex.  The
quantum terms in \fgen\ associated with the hypermultiplets, because of
the sign difference, similarly leads to a purely concave contribution
to $\cF$.  Therefore, $\cF$ will be convex provided there isn't too
much matter.  This is analogous to the requirement of asymptotic
freedom in four-dimensional gauge theories.  Finally, the
$c_{classical}$ term is neither purely convex nor purely concave on
the Weyl chamber.

We can immediately make some general comments about when there can or
cannot be a nontrivial fixed point:

\lfm{1.} There cannot be interesting fixed points
associated with Abelian gauge groups; unless there are gauge fields
which make a convex contribution to $\cF$, the matter contribution
would lead to a concave $\cF$.

\lfm{2.} On the other hand, for any non-Abelian
gauge theory without matter hypermultiplets \fgen\ is convex (taking
$c_{classical}=0$) and, therefore, there could be a fixed point.

\lfm{3.} There can be no interesting fixed point with theories
containing representations with weights ${\bf W}_f$ which are of equal
or longer length than those of the adjoint, the roots ${\bf R}$.

\lfm{4.} Turning on a quark mass $m_f$ changes the prepotential by
\eqn\deltacf{\delta \cF = {1 \over 12} \sum _{{\bf w}\in {\bf W}_f}
\left(|{\bf w}\cdot {\bf \phi}|^3-|{\bf w}\cdot {\bf \phi}+m_f|^3\right).}
We can now take $m_f \rightarrow \infty$ and derive the effective
prepotential of the problem with one fewer quark.  The terms which are
quadratic in $\phi$ can be canceled by adjusting the bare coupling
$m_0$.  The cubic terms $ {1 \over 12} \sum _{{\bf w}\in {\bf W}_f}
\left(|{\bf w}\cdot {\bf \phi}|^3-\sign (m_f)({\bf w}\cdot {\bf
\phi})^3\right)$ are
convex.  (The last term in $\delta \cF$ is the $c_{induced}$ discussed
in the last section---it vanishes for all cases except $SU(N)$, $N\geq
3$.)  Therefore, if the prepotential with some matter content is convex,
the prepotential obtained by giving mass to some matter fields and
integrating them out will also be convex---less matter means the
prepotential is even more convex.  This is consistent with flowing
{}from a nontrivial fixed point to another nontrivial fixed point upon
perturbing by mass terms.

\lfm{5.} We can immediately rule out new fixed points associated with
product gauge groups.  In order for such a theory to give something
new, the gauge groups must be coupled via matter fields which
transform nontrivially under more than one gauge group.  Concretely,
with gauge group $G_1\times G_2$, there must be hypermultiplets in
representations like $({\bf r_1}, {\bf r_2})$.  Now go along the
Coulomb branch, breaking $G_1$ without breaking $G_2$.  These matter
fields in \fgen\ lead to a negative contribution for the effective
coupling of $G_2$ which is proportional to the $G_1$ Coulomb modulus
and can thus be made arbitrarily negative.  Hence the effective gauge
coupling is not non-negative on the entire Coulomb branch and there
must be a finite bare coupling $g_2$, ruling out a fixed point.  We
thus only need to consider simple gauge groups to completely
categorize the possible fixed points.

\lfm{6.} By the decoupling of vector multiplets and hypermultiplets,
any theory obtained via Higgsing {}from a theory with convex
$\cF$ will also have a convex $\cF$.  The converse need not be true: a
theory with convex $\cF$ can be obtained via Higgsing {}from one with a
$\cF$ which is not convex---this just means that the photons
associated with the dangerous eigenvalues got lifted by the Higgsing.

In what follows we will examine in the various possible cases the
condition that the prepotential be convex, the {\it necessary}\/
conditions for a nontrivial fixed point.

\newsec{$Sp(N)$ Gauge Theories}

The Coulomb branch of the moduli space is given by $\Phi=$diag$(a_1,
\dots a_N, -a_1, \dots -a_N)$, modulo the Weyl group action, which
permutes these elements.  It can thus be taken to be the Weyl chamber
$a_1\geq a_2\geq \dots \geq a_N\geq 0$.  There are various enhanced
gauge symmetries on the walls of the Weyl chamber: with $k$ zero $a_i$
and $p$ equal but non-zero $a_i$, the unbroken gauge group is
$U(p)\times Sp(k)\times U(1)^{N-k-p}$.  We consider theories with
$n_A$ matter hypermultiplets in the antisymmetric and $n_F$ matter
flavors (that is, $2n_F$ half-hypermultiplets of matter)
in the fundamental representation of $Sp(N)$.  The global
symmetry of the classical theory is $Sp(n_A)\times Spin(2n_F)\times
U(1)_I$, where $U(1)_I$ is the symmetry \fdfp\ associated
with the current ${}^*(F\wedge F)$.

The prepotential \fgen\ with $m_0=0$ and masses $m_f=0$ is given by
\eqn\spnprep{\cF= {1\over 6}\left(\sum _{i<j}[(a_i-a_j)^3+
(a_i+a_j)^3](1-n_A)
+ \sum _ia_i^3(8-n_F)\right).}
The effective gauge coupling matrix is given by its Hessian, which is
\eqn\spmatrix{\eqalign{(g^{-2})_{ii}&=
2[(N-i)a_i+\sum _{k=1}^{i-1}a_k](1-n_A)+a_i(8-n_F),\cr
(g^{-2})_{i<j}&=2(1-n_A)a_j.}}

Note that for $n_A=1$ the matrix \spmatrix\ is diagonal and the
condition for an interacting fixed point, positivity of the diagonal
elements, is satisfied for $n_F\leq 7$.  For $n_A=1$ and $n_F=8$,
\spmatrix\ is identically zero and the Coulomb moduli space collapses
to nothing.

For arbitrary $n_A$ and $n_F$, consider the direction of the Coulomb
branch along which $Sp(N)$ is broken to $SU(k)\times Sp(N-k)\times
U(1)$: $a_1=a_2=\dots =a_k\neq 0$ with the remaining $a_i=0$.  In this
limit, the eigenvalues of
\spmatrix\ are $a_1[2(N+k-2)(1-n_A)+8-n_F]$, $k-1$ eigenvalues
$a_1[2(N-2)(1-n_A)+8-n_F]$, and $N-k$ eigenvalues $2a_1(1-n_A)$.
Therefore, we see that necessary conditions for $\cF$ to be convex and
the possibility of an interacting fixed point are: $n_A\leq 1$; for
$n_A=0$, $n_F\leq 2N+4$.  (An exception is the $Sp(1)\cong SU(2)$ case
in the original work of \fdfp, where $N=k=1$ and $g^{-2}$ is given by
the first of these eigenvalues, which is positive for $n_F\leq 7$.)

The condition, for $n_A=0$, that $n_F\leq 2N+4$ is not only a
necessary condition for convex $\cF$, and hence an interesting fixed
point, it is also sufficient.  Indeed, for $n_A=0$ and $n_F=2N+4$ the
eigenvalues of \spmatrix\ are especially simple and non-negative {\it
everywhere}\/ on the Weyl chamber; they are: $2(\sum
_{i=1}^ka_i-ka_{k+1})$, for $k=1,\dots , N-1$, and $2\sum
_{i=1}^Na_i$.  As discussed in the previous section, because $\cF$ is
convex on the entire Coulomb branch for $n_F=2N+4$, it will be convex
for any $n_F\leq 2N+4$.

Bare masses can also be introduced.  For example, giving the
fundamentals masses $m_f$, $f=1\dots n_F$, \spmatrix\ becomes
\eqn\spmatrixm{\eqalign{(g^{-2})_{ii}&=
2[(N-i)a_i+\sum _{k=1}^{i-1}a_k](1-n_A)+8a_i-\half \sum
_{f=1}^{n_F}(|a_i+m_f|+|a_i-m_f|),\cr (g^{-2})_{i<j}&=2(1-n_A)a_j.}}

The theories with $n_A=1$ and $n_F\leq 7$ are expected to have an
interacting fixed point by the same argument as in \fdfp\ but with $N$
four-brane probes.  And, as in \fdfp, the global symmetry should be
enhanced {}from $Sp(1)\times D_{n_F}\times U(1)_I$ to $Sp(1)\times
E_{n_F+1}$.  For $n_A=1$ and any $n_F$, there is a Higgs branch which,
for finite coupling, is the moduli space of $N$ $D_{n_F}$ instantons,
exactly as in \ewsmall\ in six dimensions.  For the new fixed points
with $g_{cl}^{-2}\rightarrow 0$, we expect {}from string theory that the
Higgs branch should become the moduli space of $N$ $E_{n_F+1}$
instantons.

Starting {}from the theories with $n_A=1$ and $n_F\leq 7$ and giving
the antisymmetric tensor a large mass, we can flow to the theories
with $n_A=0$ and $n_F\leq 7$.  The mass term breaks the $Sp(1)\times
E_{n_F+1}$ global symmetry of the strong coupling fixed point to
$U(1)\times E_{n_F+1}$, where the $U(1)$ acts only on the massive
antisymmetric tensor and thus decouples {}from the low-energy theory.
Therefore, at least for $n_A=0$ and $n_F\leq 7$, we expect there to be
a fixed point with $E_{n_F+1}$ global symmetry for any $Sp(N)$.  For
$n_A=0$ with $7<n_F\leq 2N+4$ our consistency condition is satisfied
but in order to construct examples we must use Calabi--Yau models.  We
shall do this in sect.~9.

As in \refs{\msdelp, \dkvdelp}, there are actually two fixed point
theories associated with the theories with $n_A=0,1$ and $n_F=0$,
associated with $\pi _4(Sp(N))=Z_2$: the analogues of $E_1$ and
$\widetilde{E}_1$ in \msdelp.  For $n_A=0$ and $n_F=0$, there is one
fixed point with global symmetry $E_1=SU(2)$ and another with global
symmetry $\widetilde E_1=U(1)$.  For $n_A=1$ and $n_F=0$, there is a
similar situation, with an additional global $Sp(1)$.  Also, exactly
as in \msdelp, there are $Sp(N)$ analogues of the $E_0$ theory, with no
global symmetry for $n_A=0$ and global symmetry $Sp(1)$ for $n_A=1$.

Consider the behavior near the boundaries of the Weyl chamber, where
$U(1)^N$ is enhanced to some bigger subgroup of $Sp(N)$.  At the
codimension $p+k-1$ boundary where $k$ of the $a_i$ are zero and $p$
equal but non-zero, there is an enhanced $U(p)\times Sp(k)\times
U(1)^{N-k-p}$ with ${\bf 2N}\rightarrow ({\bf p}_{\pm 1}, {\bf 1})_{(0,
\dots ,0)}+({\bf 1}_0, {\bf 2k})_{(0, \dots , 0)}+(({\bf 1}_0, {\bf
1})_{\pm 1, 0, \dots 0}$+permutations).  Similarly decomposing the
$Sp(N)$ vector multiplet and taking into account the masses that
fields charged under the $U(1)s$ get along this direction of the
Coulomb branch, the prepotential \spnprep\ is, of course, properly
reproduced in the low-energy theory.

For $n_F\geq 2$ there is a Higgs branch, which connects to the Coulomb
branch along the boundaries of the Weyl chamber discussed above with
$k>0$.  In particular, along $a_N=0$, $Sp(N)$ with $n_F$ flavors can
be Higgsed to $Sp(N-1)$ with $n_F-2$ flavors.  Setting $a_N=0$ in
\spnprep, the prepotential reduces to exactly that of $Sp(N-1)$ with
$n_F-2$ flavors.  This is in agreement with the expected decoupling of
vector and hypermultiplets.

It is interesting to compare our condition for an interacting fixed
point in five dimensions to the condition of asymptotic freedom in
four dimensions.  Reducing the five-dimensional supersymmetric theory
to four dimensions yields a theory with $N=2$ supersymmetry.  There
the condition for asymptotic freedom for $Sp(N)$ with $n_A$
hypermultiplets in the antisymmetric representation and $n_F$ in the
fundamental is $2(N+1)-n_F-2n_A(N-1)\geq 0$.  For example, for $n_A=0$
the condition is $n_F\leq 2(N+1)$, which is stronger than our
condition in five dimensions.  On the other hand, for $N=2$ and
$n_F=0$, there can be $n_A\leq 3$ for four-dimensional asymptotic
freedom, which is weaker than our condition in five dimensions.  So,
in general, the condition for a fixed point in five dimensions can be
either stronger or weaker than that of asymptotic freedom in four
dimensions.

\subsec{$Sp(2)\cong Spin(5)$ in more detail}

The Weyl chamber is given by $a_1\geq a_2\geq 0$.  On the boundary $a_2=0$
there is an unbroken $Sp(1)\times U(1)$ with the fundamental decomposing
as ${\bf 4}\rightarrow {\bf 2}_0+{\bf 1}_{\pm 1}$, the antisymmetric
tensor decomposing as ${\bf 5}\rightarrow {\bf 2}_{\pm 1}+{\bf 1}_0$,
and the adjoint decomposing as ${\bf 10}\rightarrow {\bf 3}_0+{\bf
2}_{\pm 1}+{\bf 1}_{\pm 2}+{\bf 1}_0$.  Along this boundary, the $SU(2)$
has $n_F$ massless doublets, with those coming {}from the ${\bf 5}$ for
$n_A=1$ getting a mass {}from their coupling to the $U(1)$.  On the
boundary $a_1=a_2$ there is an unbroken $SU(2)\times U(1)$, with the
representations decomposing as ${\bf 4}\rightarrow {\bf 2}_{\pm 1}$,
${\bf 5}\rightarrow {\bf 3}_0+{\bf 1}_{\pm 2}$, ${\bf 10}\rightarrow
{\bf 3}_{\pm 2}+{\bf 3}_0+{\bf 1}_0$.  The unbroken $SU(2)$, with zero
bare masses, has no massless doublets, as those coming {}from the ${\bf
4}$'s get a mass via their coupling to the $U(1)$.  For $n_A=1$, however,
there can be a massless ${\bf 3}$ of $SU(2)$.

By introducing a bare mass $m$ for the matter
fields in the ${\bf 4}$, the unbroken $SU(2)$ will have $n_F$ massless
doublets at $a_1=a_2=|m|$.  On the boundary $a_1=a_2$, on one side of
this point, the $SU(2)$ will go to the $E_1$ fixed point, and on the
other side it will flow to the $\widetilde E_1$ fixed point discussed
in \msdelp.

For $n_F\geq 2$ and zero bare masses, there is a Higgs branch which
meets the Coulomb branch at the boundary $a_2=0$.

Theories with $n_A=1$ and $n_F\leq 7$ lead to new fixed points via a
generalization of the argument in \fdfp\ to one involving two probes.
By giving a mass to the ${\bf 5}$ and integrating it out, the theories
with $n_A=0$ and $n_F\leq 7$ thus also lead to new fixed points.  By
the general analysis above, there is one more case where there could
possibly be an interesting fixed point: $n_A=0$, $n_F=8$.  The
existence of this fixed point, and more generally all of the possible
fixed points with $n_F\leq 2N+4$, is demonstrated in sect.~9.

\newsec{$SU(N)$ Gauge Theories}

The Coulomb branch of the moduli space is given by $\Phi=$diag$(a_1,
\dots a_N)$, with $\sum _ia_i=0$, modulo the Weyl group action, which
permutes the $a_i$.  It can thus be taken to be the Weyl chamber
$a_1\geq a_2\geq \dots \geq a_N$.  On the walls of the Weyl chamber
there is enhanced gauge symmetry: with $k$ equal $a_i$, $U(1)^{N-1}$
is enhanced to $U(k)\times U(1)^{N-1-k}$.

We only need to consider theories with matter in the symmetric,
antisymmetric, and fundamental representations; $n_S$, $n_A$, and
$n_F$ are the number of hypermultiplets in these representations.  For
other representations, there cannot be an interesting fixed point.
(We will shortly argue that $n_S=0$ is also necessary.)  There is a
$U(n_S)\times U(n_A)\times U(n_F)\times U(1)_I$ classical global
symmetry.  As discussed in sect.~2.2, it is possible and sometimes
necessary to include a classical $c_{cl}$ term for $SU(N)$.  The
condition on $c_{cl}$ is
\eqn\Ccond{c_{cl}+\half n_F+\half N(n_A+n_S)\in \IZ,}
where we used the fact that the (anti)symmetric tensor representation
has cubic index $N+(-)4$.

The effective prepotential on the Coulomb branch, taking $m_0=0$, is
given by (the case with no matter and $c_{cl}=0$ has already appeared
in \ref\nnek{N. Nekrasov, ``Five Dimensional Gauge Theories and
Relativistic Integrable Systems,'' hep-th/9609219.})
\eqn\sunprep{\cF={1\over 12}\left(2\sum ^N_{i<j}(a_i-a_j)^3+
2c_{cl}\sum _{i=1}^Na_i^3-(n_A+n_S)\sum
_{i<j}^N|a_i+a_j|^3-(n_F+8n_S)\sum _{i=1}^N|a_i|^3\right).}  The
prepotential \sunprep\ should be written in terms of $N-1$ independent
variables, solving for the constraint $\sum _{i=1}^Na_i=0$.  Because a
symmetric tensor makes the same contribution to $\CF$ as an
antisymmetric tensor and $8$ fundamentals, we will no longer need to
write the dependence on $n_S$ explicitly.

With $n_F,n_A,n_S\neq 0$, there is not a single prepotential but, as
discussed in sect.~2.3, prepotentials defined in sub-wedges of the
Weyl chamber corresponding to different values of the signs of those
terms in \sunprep\ which have absolute values.  There are massless
charged matter fields at the boundaries of these sub-wedges.  For
example, for $n_A+n_S=0$ there are $N-1$ sub-wedges in the Weyl
chamber corresponding to the $N-1$ choices of where zero appears in
$a_1\geq a_2\geq \dots
\geq a_N$; in the $k$-th sub-wedge, $k$ of the $a_i$ are negative.
(For $n_F\geq 2$ there is a Higgs branch which connects to the Coulomb
branch along the boundaries of these sub-wedges.)  The prepotential in
the $p$-th sub-wedge, for $p=2\dots N-1$, is given by $\cF _{(p)}=\cF
_{(1)}+(n_F/6)\sum _{i=1}^{p-1}a_{N-i}^3$, where $\cF _{(1)}$ is the
prepotential for the first sub-wedge, where $a_{N-1}\geq 0$.

We now find the conditions on the number of flavors and $c_{cl}$ for
the prepotential \sunprep\ to be convex throughout the Weyl chamber.
Because the charge conjugation operation, which acts on the Weyl
chamber as $a_i\rightarrow -a_{N+1-i}$, $i=1\dots N$, takes
$c_{cl}\rightarrow -c_{cl}$, any condition on $c_{cl}$ will imply a
similar condition on $-c_{cl}$.  Therefore, the convexity condition on
$c_{cl}$ will be in terms of $|c_{cl}|$.

In the sub-wedge with $a_{N-1}\geq 0$, the metric $g^{-2}$ in terms of
the Hessian of $\cF$ with respect to the $a_i$, $i=1\dots N-1$, with
$a_N=-a_T\equiv -\sum _{i=1}^{N-1}a_i$, is given by:
\eqn\sumatrix{\eqalign{(g^{-2})_{ii}={}
&(N+4+c_{cl}-2i)a_i+2\sum _{k=1}^{i-1}a_k \cr& +(N+2-c_{cl})a_T-\half
((N-2)n_A+n_F)(a_i+a_T),\cr (g^{-2})_{i<j}={}&(2-n_A)a_j+\half
(2N+4-n_A(N-4)-n_F-2c_{cl})a_T-n_Aa_i.}}  The metric $g^{-2}$ in the
$p$-th sub-wedge differs {}from \sumatrix\ by $\Delta ^{(p)}_{ij}=\sum
_{k=1}^{p-1}\delta _{N-k,i}\delta _{ij}$.

Along the direction where $a_i=a$, $i=1\dots N-1$, where $SU(N)$ is
broken to $SU(N-\nobreak1)\times U(1)$, \sumatrix\ has $N-2$
eigenvalues $a[2N-n_A(3N-8)-n_F+2c_{cl}]/2$ and one eigenvalue
$a[2N^3-n_A(N-2)(N^2-4N+8)-n_F(N^2-2N+2)-2c_{cl}N(N-2)]/2$.  For $N>2$
a necessary condition for convex prepotential is thus
$2N-n_A(3N-8)-n_F-2|c_{cl}|\geq 0$, where the absolute value follows
{}from the charge conjugation operation mentioned above.  When this
condition is satisfied, both of the above eigenvalues are
non-negative.  Putting $n_A=n_S$ and $n_F=8n_S$, we see that $n_S=0$
is required for the first eigenvalue to be non-negative.  Similarly,
$n_A=0$ is required for $N>8$.  For $N\leq 8$, these eigenvalues are
also non-negative with $n_A=1$ antisymmetric tensor and $n_F\leq
8-N-2|c_{cl}|$ fundamentals.  For $N=4$ they are also non-negative for
$n_A=2$, $n_F=c_{cl}=0$.  For all $N$, with $n_A=0$, we get
\eqn\nfcmax{n_F+2|c_{cl}|\leq 2N}
as a necessary condition for a nontrivial fixed point.
Thus $n_F\leq 2N$ and $|c_{cl}|\leq N$.

For $n_F=2N$, \nfcmax\ requires $c_{cl}=0$, which is compatible with
\Ccond.  For $n_F=2N$ and $c_{cl}=0$ the gauge coupling matrix
\sumatrix\ has especially simple eigenvalues: $2(\sum
_{i=1}^ka_i-ka_{k+1})$, for $i=1, \dots , N-2$, and $2N\sum
_{i=1}^{N-1}a_i$.  Because these are all non-negative in the Weyl
chamber, $n_F\leq 2N$ is, in fact, necessary and sufficient for having
$\cF$ be convex in the entire first sub-wedge.  The prepotential in
the $p$-th sub-wedge, for $p=2\dots N-1$, is given by $\cF _{(p)}=\cF
_{(1)}+(n_F/6)\sum _{i=1}^{p-1}a_{N-i}^3$, where $\cF _{(1)}$ is the
prepotential for the first sub-wedge, which we just found to be convex
throughout the entire Weyl chamber.  The additions in the ${(p)}$
sub-wedge, on the other hand, are purely concave since the
$a_{N-i}<0$.  Therefore, one might worry that $\cF _{(p)}$ could fail
to be convex.  But even in the $N-1$-th sub-wedge, which is the worst
case in terms of possibly not being convex, $\cF$ is, in fact, convex.
Indeed, the $N-1$-th sub-wedge is related to the first sub-wedge by
taking $\Phi \rightarrow -\Phi$ (for $c_{cl}=0$ this reflects the
underlying charge conjugation symmetry), {}from which it follows that
$\cF _{(N-1)}$ will be convex since $\cF _{(1)}$ is.

Since the prepotential is convex for zero quark masses for $n_F=2N$
and $c_{cl}=0$, it follows {}from comment 4 in sect.~3 that it is also
convex when the masses are turned on.  It also follows that the
inequality
\nfcmax\ is necessary and sufficient for all $n_F$ and $c_{cl}$.
Any theory satisfying \nfcmax\ can be obtained {}from $n_F=2N$ and
$c_{cl}=0$ by adding mass terms.  Different values of $c_{cl}$ are
induced by choosing appropriate signs for the mass terms (see equation
\effec).  By comment 4, they will all have convex prepotentials.
Therefore, there can be interesting fixed points whenever \Ccond\ and
\nfcmax\ are satisfied.  An exception is $SU(2)$, where $c_{cl}$ doesn't
arise and where the fixed points extend to $n_F\leq 7$ \fdfp.

For $n_F\geq 2$ there is a Higgs branch associated with the Higgsing:
$N\rightarrow N-1$, $n_F\rightarrow n_F-2+n_A$, $n_A\rightarrow n_A$.
This Higgs branch connects to the Coulomb branch along the $N-2$
boundaries of the sub-wedges, where $a_l=0$ for $l=2\dots N-2$.  The
above expressions are compatible with this Higgsing.  In particular,
setting $a_l=0$ in \sunprep, the prepotential exactly reproduces that
of the low-energy $SU(N-1)$ theory obtained along the Higgs branch.

Bare masses can be added for the matter fields, with the prepotential
modified as in \fgen.  Giving the fundamentals masses $m_f$, this
modifies the metric \sumatrix\ by replacing $n_Fa_i$ with $\sum
_{f=1}^{n_F}|a_i+m_f|$ and $n_Fa_T$ with $\sum _{f=1}^{n_F}|a_T-m_f|$.
The mass dependence is compatible with flowing between fixed points
upon adding masses for the matter fields, with an addition of
$c_{induced}$ as in \inducedc\ and a shift of $m_0$.

\subsec{$SU(3)$ in more detail}

We now consider $SU(3)$ in a bit more detail, explicitly verifying
that the gauge coupling eigenvalues are indeed non-negative for all
$n_F\leq 6$.  Taking $a_3=-a_1-a_2$, the Weyl chamber is $a_1\geq
a_2\geq -\half a_1$.  At this boundary of the Weyl chamber there is an
unbroken $SU(2)\times U(1)$ with the fundamental decomposing as ${\bf
3}\rightarrow {\bf 2}_1+{\bf 1}_{-2}$ and the adjoint as ${\bf
8}\rightarrow {\bf 3}_0+{\bf 2}_{\pm 3}+{\bf 1}_0$.  Because of the
non-zero $U(1)$ charge of the matter in the ${\bf 2}_1$, the $SU(2)$
theory has no massless matter fields for zero bare mass.  By
introducing a bare mass $m$ for the matter fields, the point
$a_1=a_2=|m|$ has an $SU(2)$ with $n_F$ massless doublets.  On the
rest of the Weyl boundary, the doublets get a mass of one sign on one
side of this point and another sign on the other side of this point.

For $n_F>0$ the prepotential \sunprep\ differs in the two sub-wedges,
$a_2\geq 0$ and $a_2\leq 0$, of the Weyl chamber.  Explicitly, the
metric in the sub-wedge $a_1\geq a_2\geq 0$ is
\eqn\suiiimeti{t(a)=\pmatrix{(10-n_f)a_1+(5-c_{cl}-\half n_f)a_2&
2a_2+(5-c_{cl}-\half n_f)(a_1+a_2)\cr 2a_2+(5-c_{cl}-\half n_f)(a_1+a_2)
& (7-c_{cl}-\half n_f)a_1+(8-n_f)a_2}.}
In the sub-wedge $0\geq a_2\geq -\half a_1$ the metric is given by
\eqn\suiiimetii{t(a)=\pmatrix{(10-n_f)a_1+(5-c_{cl}-\half n_f)a_2&
2a_2+(5-c_{cl}-\half n_f)(a_1+a_2)\cr 2a_2+(5-c_{cl}-\half n_f)(a_1+a_2)
& (7-c_{cl}-\half n_f)a_1+8a_2}.}
The eigenvalues of these are complicated in general but are
non-negative when \nfcmax\ is satisfied.

\newsec{$Spin(M)$ Gauge Theories}

For $Spin(2N)$ or $Spin(2N+1)$ the Coulomb branch moduli space is
given by the Weyl chamber $a_1\geq a_2\geq \dots \geq a_N\geq 0$,
corresponding to the block diagonal elements of $\Phi$.  On the
boundary of the Weyl chamber with $k$ zero $a_i$'s and $p$ equal but
non-zero $a_i$'s, there is an unbroken $Spin(2k+\epsilon)\times
U(p)\times U(1)^{N-k-p}$ gauge group, where $\epsilon =0$ $(1)$ for
$M$ even (odd).

Because the antisymmetric tensor representation is the adjoint, there
cannot be any matter fields in the symmetric or antisymmetric tensor
representations.  The theory with $n_V$ hypermultiplets in the vector
representation has a classical global $Sp(n_V)\times U(1)_I$ symmetry.
Our discussion will be for $Spin(M)$ with $M\geq 5$; $Spin(3)\cong SU(2)$
is covered by the original work of
\fdfp\ and, as discussed in sect.~3, $Spin(4)\cong SU(2)\times
SU(2)$ is not interesting because it is not simple.

The prepotential for $Spin(2N)$ with $n_V$ hypermultiplets in the vector
representation is
\eqn\soneprep{\cF={1\over 6}\left(\sum
_{i<j}[(a_i+a_j)^3+(a_i-a_j)^3]-n_V\sum _ia_i^3\right).}
The corresponding effective gauge coupling matrix is
\eqn\sonematrix{\eqalign{(g^{-2})_{ii}&=
2[(N-i)a_i+\sum _{k=1}^{i-1}a_k]-n_Va_i,\cr
(g^{-2})_{i<j}&=2a_j.}}
The prepotential for $Spin(2N+1)$ is
\eqn\sonoprep{\cF={1\over 6}\left(\sum
_{i<j}[(a_i+a_j)^3+(a_i-a_j)^3]+(1-n_V)\sum _ia_i^3\right).}
The corresponding matrix is
\eqn\sonomatrix{\eqalign{(g^{-2})_{ii}&=
2[(N-i)a_i+\sum _{k=1}^{i-1}a_k]+(1-n_V)a_i,\cr
(g^{-2})_{i<j}&=2a_j.}}

Consider the direction in the Coulomb branch where $Spin(M)$ is broken
to $U(k)\times Spin(M-2k)$: $a_1=\dots =a_k$, with the remaining
$a_i=0$.  In this limit, \sonematrix\ or \sonomatrix\ has an
eigenvalue $a_1[M+2k-4-n_V]$, $k-1$ eigenvalues $a_1(M-4-n_V)$, and
$N-k$ eigenvalues $2a_1k$.  Therefore, a necessary condition for
non-negative eigenvalues is $n_V\leq M-4$.  For $n_V=M-4$ the
eigenvalues of \sonematrix\ and \sonomatrix\ simplify; they are
$2(\sum _{i=1}^pa_i-pa_{p+1})$, $p=1\dots N-1$, and $2\sum
_{i=1}^Na_i$.  These eigenvalues are indeed all non-negative
throughout the Weyl chamber, showing that $n_V\leq M-4$ is a necessary
and sufficient condition for everywhere convex prepotential $\cF$.
(The condition for asymptotic freedom in the four-dimensional, $N=2$
version of this is $n_V\leq M-2$.)

In the direction on the Coulomb branch where there are $k$ equal
eigenvalues $a_i=a$, $Spin(M)$ is broken to $SU(k)\times U(1)^{N-k+1}$.
If the $n_V$ matter fields originally had masses $m_i$, the $SU(k)$
theory will have $n_F=2n_V$ matter fields with pairs having masses
$m_i\pm a$.

For $n_V\geq 1$ there is a Higgs branch corresponding to the Higgsing:
$M\rightarrow M-1$ with $n_V\rightarrow n_V-1$.  For $M=2N$, this
Higgs branch connects to the Coulomb branch along the boundary $a_N=0$
of the Weyl Chamber.  For $M=2N+1$ this Higgs branch connects to the
entire Coulomb branch.  The above expressions are compatible with this
Higgsing.  In particular, starting {}from $Spin(2N)$, setting $a_N=0$
in \soneprep\ the prepotential exactly reduces to that of $Spin(2N-1)$
with $n_V-1$ vectors.  Similarly, starting {}from $Spin(2N+1)$,
\sonoprep\ is, over the entire Coulomb branch, exactly the same as the
prepotential \soneprep\ for the $Spin(2N)$ theory with $n_V-1$ vectors
obtained along the Higgs branch.

The dimension of the full Higgs branch for $n_V\leq M$ is
$\half n_V(n_V+1)$, with the gauge group generically broken to
$Spin(M-n_V)$ with no matter.  In particular, for our critical case of
$n_V=M-4$, the gauge group can be broken to $Spin(4)\cong SU(2)\times
SU(2)$ with no matter on the Higgs branch.  A simple way to see that
$n_V\leq M-4$ is necessary for a new fixed point is to note that for
larger $n_V$ it is possible to Higgs to $Spin(4)$ with matter fields
in the $({\bf 2}, {\bf 2})$ which, as discussed in sect.~3,
necessarily leads to negative eigenvalues.

Note that for $n_V=M-4$ or $n_V=M-3$ the Higgs branch connects to
$Spin(5)\cong Sp(2)$ with $n_A=0$ or $n_A=1$ antisymmetric tensors
which, as discussed in sect.~2, we know have nontrivial fixed points
in analogy with $E_1$, $\widetilde E_1$, and $E_0$ of \msdelp.  For
$n_V=0$, for $M>5$, there is expected to be only one fixed point
because $\pi _4(Spin(M))=0$.

We could include $n_S$ spinors by including a term in the prepotential
\soneprep\ for $M=2N$:
$$-{1\over 12} n_S\sum_{\{\epsilon _i\}} |\half \sum _i\epsilon
_ia_i|^3,$$ where $\{\epsilon _i\}$ runs over all $2^{N-1}$ sign
choices for $\epsilon _i=\pm 1$ with an even (odd) number of $-$ signs
for the spinor (conjugate spinor) representation.  In the limit where
only $a_1\neq 0$, this would lead to an eigenvalue
$a_1[2(N-1)-n_V-n_S2^{N-5}]$ and $N-1$ eigenvalues
$2a_1(1-n_S2^{N-6}),$ showing that we need, for any $n_V$, $n_S\leq
2^{6-N}$ and thus $n_S=0$ for $M=2N>12$.  Considering also the limit
where only $a_1=a_2\neq 0$ again yields $n_V\leq M-4$ as a necessary
condition on $n_V$ for any $n_S$.  Note that when $n_V=M-4$ there is a
Higgs branch where $Spin(2N)$ is broken to $Spin(4)\cong SU(2)\times
SU(2)$ with $2^{N-3}n_S$ hypermultiplets in the $({\bf 2}, {\bf
1})+({\bf 1}, {\bf 2})$.  The condition that $n_S\leq 2^{6-N}$ then
corresponds to the condition that each $SU(2)$ can have at most $8$
flavors before leading to negative eigenvalues.  Similar
considerations apply for $M=2N+1$, showing that $n_S\leq 2^{5-N}$.

\newsec{Exceptional Gauge Groups}

\subsec{$G_2$}

$G_2$ with $n_7$ matter fields in the ${\bf 7}$ has a $Sp(n_7)\times
U(1)_I$ global symmetry.  We can obtain the conditions for $G_2$
simply by decomposing it in terms of $SU(3)$: ${\bf 7}\rightarrow {\bf
3}+\overline {\bf 3}+{\bf 1}$, ${\bf 14}\rightarrow {\bf 8}+{\bf
3}+{\bf \overline 3}$.  Therefore, the effective gauge coupling for
$G_2$ can be obtained by substituting $n_3=2(n_7-1)$ in the $SU(3)$
answer.  Corresponding to the condition $n_3\leq 6$ found in sect.~3,
the necessary condition for convex $\cF$ for $G_2$ is thus $n_7\leq
4$.

In terms of the variables $a_i$ used in sect.~3 for
$SU(3)$, the Weyl chamber for $G_2$ is $a_1\geq a_2\geq 0$.  At the
boundary $a_1=a_2$, $G_2$ is broken to $SU(2)\times U(1)$, with the
representations decomposing as ${\bf 7}\rightarrow {\bf 2}_{\pm
1}+{\bf 1}_{\pm 2}+{\bf 1}_0$ and ${\bf 14}\rightarrow {\bf 2}_{\pm 1}+
{\bf 1}_{\pm 2}+ {\bf 3}_0+{\bf 2}_{\pm 3}+{\bf 1}_0$.  At the other
boundary, $a_2=0$, $G_2$ is broken to $SU(2)'\times U(1)$, with
the representations decomposing as ${\bf 7}\rightarrow {\bf 3}_0+{\bf
2}_{\pm 1}$ and ${\bf 14}\rightarrow {\bf 3}_0+{\bf 4}_{\pm 1}+{\bf
1}_{\pm 2}+{\bf 1}_0$.  For $n_7>0$ there is everywhere a Higgs
branch with $G_2$ broken to $SU(3)$ according to the above
decomposition.

\subsec{$F_4$}

The classical theory with $n_{26}$ matter fields in the fundamental
representation has an $Sp(n_{26})\times U(1)_I$ global symmetry.  To
find the condition for a nontrivial fixed point, we can decompose in
terms of $Spin(9)$: ${\bf 26}\rightarrow {\bf 9}+{\bf 16}+{\bf 1}$ and
${\bf 52}\rightarrow {\bf 36}+{\bf 16}$.  Therefore, the effective
gauge coupling for $F_4$ can be obtained by substituting $n_V=n_{26}$
and $n_S=n_{26}-1$ in the $Spin(9)$ answer.  This gives $n_{26}\leq 3$
as a necessary (though perhaps not sufficient) condition for convex
$\cF$.  For example, with $n_{26}=3$ there is a Higgs branch to
$Spin(6)\cong SU(4)$ with $8$ matter fields in the ${\bf 4}$, just
making the condition for non-negative eigenvalues.  On the other hand,
for $n_{26}=4$ there is a Higgs branch to $Spin(5)\cong Sp(2)$ with $12$
matter fields in the ${\bf 4}$, which yields a negative eigenvalue.

\subsec{$E_6$, $E_7$, and $E_8$}

For $E_6$ there could be a nontrivial fixed point associated with
theories with $n_{27}$ matter fields in the ${\bf 27}$.  The classical
global symmetry is $SU(n_{27})\times U(1)_I$.  For $n_{27}>0$,
there is a Higgs branch with $E_6$ broken to $F_4$ with
$n_{26}=n_{27}-1$.  Corresponding to the necessary (though perhaps not
sufficient) condition found for $F_4$, a necessary (though perhaps not
sufficient) condition for convex $\cF$ is $n_{27}\leq 4$.

For $E_7$ there could be nontrivial fixed points with $n_{56}$ matter
fields in the ${\bf 56}$.  Because the ${\bf 56}$ is pseudo-real,
$n_{56}$ can be half-integral. Unlike the $Sp(N)$ case, where an
anomaly required an even number of half-hypermultiplets, here there is
no such anomaly and the number $n_{56}$ can be
half-integral\foot{Indeed, compactification of the heterotic string on
$K3\times S^1$ leads to $E_7$ in five dimensions with $n_{56}=\half
k-2$ for an $E_8$ with $k$ instantons.}.  The classical global
symmetry is $Spin(2n_{56})$.  For $n_{56}\geq 0$, there is a Higgs
branch with $E_7$ broken to $E_6$ with $n_{27}=2(n_{56}-1)$.
Corresponding to the condition found above for $E_6$, $n_{56}\leq 3$
is thus a necessary (though perhaps not sufficient) condition for
convex $\cF$.

For $E_8$, because the fundamental representation is the adjoint,
there can only be a new fixed point for the theory without matter.
Because $\pi _4(E_8)=0$, there should only be one such theory.

\newsec{Calabi--Yau Interpretation}

We now turn to the Calabi--Yau interpretation of these theories.
Non-Abelian gauge symmetry in stringy models compactified on a
Calabi--Yau threefold arises {}from singularities along holomorphic
curves in the threefold \refs{\BSV,\aspinwall}.  This happens in type
IIA compactifications, M-theory compactifications, and F-theory
compactifications (to four, five and six dimensions, respectively),
with the same geometry governing all cases
\refs{\KM,\KMP,\mandf,\MVi}.  The easiest way to describe the geometry
is by considering a desingularization $\pi:X\to\overline{X}$ of the
original Calabi--Yau space $\overline{X}$, by a Calabi--Yau manifold
$X$ which contains a collection of complex surfaces $S_j$ that shrink
to a holomorphic curve $\overline{C}$ on $\overline{X}$.  (This
``shrinking'' is accomplished by sending the K\"ahler class of $X$ to
a point on an appropriate boundary wall of the K\"ahler cone.)  There
are holomorphic curves $\sigma_\alpha$ within those surfaces which
shrink to zero size in this limit, and map to points on
$\overline{X}$; among these are the generic fiber $\varepsilon_j$ of
the map $S_j\to\overline{C}$, for each $j$.

The gauge group for M-theory compactified on $X$ is\foot{For the type
IIA compactification on $X$, the gauge group is $(H^0(X,\IR)\oplus
H^2(X,\IR))/(H^0(X,\IZ)\oplus H^2(X,\IZ))$.  If $X$ is elliptically
fibered, then the Cartan subgroup of the gauge group for the
$F$-theory compactification on the base of the fibration is
$H^2(X,\IR)_0/H^2(X,\IZ)_0$, where $H^2(X)_0$ is the subspace of
$H^2(X)$ orthogonal to the class of the elliptic curve.  The
non-Abelian enhancement in these cases is similar to that of the
M-theory compactification.}  $H^2(X,\IR)/H^2(X,\IZ)$.  On
$\overline{X}$, this is enhanced to a non-Abelian group whose simple
roots correspond to the cohomology classes $[S_j]\in H^2(X,\IZ)$.  The
charged matter which becomes massless when the symmetry is enhanced
arises {}from $2$-branes wrapping connected holomorphic curves $\sigma$
on $X$ which shrink to points in $\overline{X}$.  (More precisely,
each such curve determines a hypermultiplet which contains $2$-branes
wrapping both the holomorphic curve $\sigma$ and the anti-holomorphic
curve $\overline{\sigma}$.) The homology classes $[\sigma]\in
H_2(X,\IZ)$ and $[\overline{\sigma}]=-[\sigma]$ determine the charges,
so $H_2(X,\IZ)$ should be identified with the weight lattice.  The
homology classes of the generic fibers $\varepsilon_j$ are the adjoint
weights $[\varepsilon_j]$ dual to the simple roots $[S_j]$.

The geometric configurations of surfaces $S_j$ which give rise to all
possible non-Abelian gauge groups are known.  Given such a collection
of surfaces, the charged matter spectrum is determined by the
parameter curves for the fibers $\varepsilon_j$ and by the possible
singular fibers \refs{\AG,\rKMnew}.  In many cases, it is known how to
relate specific types of singular fibers to specific matter
representations
\refs{\MVii\BKKM\BIKMSV\sadov{--}\KV}.  Here, we run the arguments in
reverse, and determine the geometric structure when the matter
representation is given, using the fact that for each class $[\sigma]$
in the matter representation, $\pm[\sigma]$ must be represented by a
connected holomorphic curve.

For the compactification of M-theory on a Calabi--Yau threefold $X$,
the scalars in the vector multiplets parameterize the K\"ahler classes
of unit volume, and the prepotential is determined by the intersection
numbers among those classes \refs{\ccdf,\fkm,\fms}.  Concretely, if we
choose a basis $J_i$ for the K\"ahler classes and write an
arbitrary class in the form $J=\sum \phi ^iJ_i$, then these
scalars can be identified with the $\phi ^i$'s (subject to the
constraint of the volume being one), and the coefficients $c_{ijk}$ in
$\cF={1\over 6}\sum _{ijk}c_{ijk}\phi ^i\phi ^j\phi ^k$ are the
topological intersection numbers $c_{ijk}=\int J_i\wedge J_j\wedge J_k$.

The K\"ahler classes on $X$ are related to the positive classes on
$\overline{X}$ as follows.  Given a positive class $\overline{\kappa}$ on
$\overline{X}$, consider the {\it relative K\"ahler cone}
\eqn\relkk{
\cK(X/\overline{X})=\{S=\sum\psi_jS_j\ |\
\overline{\kappa}+S \hbox{ defines
a K\"ahler class on $X$}\}.  } According to the Kleiman criterion for
ampleness \kleiman, this cone can also be described as
\eqn\relkkbis{
\cK(X/\overline{X})=\{S=\sum\psi_jS_j\ |\ S\cdot\sigma>0 \hbox{ for all
$\sigma$ mapping to points on $\overline{X}$}\}, } which shows that
the definition is independent of $\overline{\kappa}$.  This
description can be used in two ways: the holomorphic classes $\sigma$
can be used to characterize the cone, or the cone can be used to
determine which of $\pm\sigma$ is holomorphic.

The mathematical structure of the relative K\"ahler cone is known in
considerable detail \kawamata.  For many purposes, it is better to
work with the negative of this cone, defined by
\eqn\relkkminus{
-\cK(X/\overline{X})=\{S=\sum\psi_jS_j\ |\ -S\cdot\sigma>0 \hbox{ for
all $\sigma$ mapping to points on $\overline{X}$}\}; } this has the
advantage that the coefficients $\psi_j$ will be positive.  We
identify the set of volume one elements in this cone as being the
portion of the vector multiplet moduli space which is relevant for
discussing the contraction $X\to\overline{X}$.

It is possible for $\overline{X}$ to have more than one
desingularization which is a Calabi--Yau manifold.  That is, there
could be distinct Calabi--Yau manifolds $X_1$, $X_2$, \dots, $X_k$ all
having maps $\pi_\alpha:X_\alpha\to\overline{X}$ which resolve the
singularities.  In this situation, the various Calabi--Yau manifolds
must differ by flops \refs{\kawamata,\flops}.  There are canonical
identifications among the spaces $H^2(X_\alpha,\IR)$, in which the
K\"ahler cones of the various $X_\alpha$'s share common boundary
walls.  The vector multiplet moduli space should include the set of
volume one elements within the union of all of these cones
\refs{\AGM,\mandf}.

The flops will affect the relative K\"ahler cones in a similar way:
the cones $-\cK(X_\alpha/\overline{X})$ will share common boundary
walls, and the relevant part of the vector multiplet moduli space
should include the set of volume one elements within the union of all
of these cones.

The prepotential in these theories is determined by the intersection
properties of divisors on $X_\alpha$.  Specifically, if we write a
general element in the negative of the relative K\"ahler cone
$-\cK(X_\alpha/\overline{X})$ in the form $S=\sum\psi_j S_j$, then the
coefficients $\psi_j$ are scalars which parameterize the moduli space,
in terms of which the prepotential can be written
\eqn\CYpp{
\cF={1\over6}\sum _{ijk}c_{ijk}\psi _i\psi _j\psi _k,
} with the coefficients $c_{ijk}$ coinciding with the intersection
numbers $S_i\cdot S_j\cdot S_k$, calculated on the threefold
$X_\alpha$.  (We can express this rather compactly by writing
$\cF={1\over6}S^3$.)  Since we are using a basis of K\"ahler classes
that came {}from integral cohomology classes $S_j\in
H^2[X_\alpha,\IZ)$, the coefficients $c_{ijk}$ are integers, as
expected {}from the quantization condition \abcijk.  Notice that when
we move {}from $X_\alpha$ to $X_\beta$, these coefficients will change.

In the remainder of this section, we discuss the detailed geometry on
the Calabi--Yau space corresponding to various specific groups, with
specified matter content.  We include adjoint matter in each of the
cases we discuss (even though it is not needed for strong coupling
limits), since the geometry associated to the adjoint matter is
somewhat intertwined with that of the other matter representations.
The number of hypermultiplets in the adjoint representation is
determined by the genus of the parameter curves for the rulings on the
ruled surfaces \KMP.

Our primary goal will be to make the geometry sufficiently explicit
that the prepotential $\cF={1\over6}S^3$ can be calculated in detail.
We will find that in each case we consider, there is a remarkable
correspondence with the group-theoretic calculations made earlier.
The combinatorial manipulations which are needed to fully check this
are quite involved, and the agreement we obtain looks like another
``string miracle,'' as anticipated by Harvey and Moore \jhgm.

We limit our calculations to the classical groups, with some
restriction on the matter content beyond that suggested by the
strong-coupling limit problem.  (These restrictions simplify the
geometry substantially.)  We are confident that similar calculations
with more general matter content or for the exceptional groups would
be equally successful (as was suggested explicitly for $E_8$ in
\jhgm).  We will turn to the consideration of strong coupling limits
in the next section.

\subsec{$Sp(N)$}

We begin with the case of gauge group $Sp(N)$, with hypermultiplets in
$g$ copies of the adjoint representation, $n_A$ copies of the
antisymmetric representation, and $n_F$ copies of the fundamental
representation. (More precisely, we have $2n_F$ half-hypermultiplets
in the fundamental representation.)  In order to produce $Sp(N)$ gauge
symmetry when the ruled surfaces $S_j$ shrink to zero size, one of
them---labeled $S_N$ below---must be ruled over a holomorphic curve of
genus $g$, with the others being ruled over a holomorphic curve of
genus $g'$ which has a two-to-one map to the curve of genus $g$.
(This corresponds to the fact that the first $N-1$ roots of $Sp(N)$
have length half that of the $N$-th root.)  The surfaces $S_j$ and
$S_{j+1}$ must meet along a holomorphic curve $\gamma_j$.  When
$j<N-1$ this curve is a section for the ruling on each surface, while
$\gamma_{N-1}$ is a section on $S_{N-1}$ and a $2$-section on $S_N$.
All of these curves of intersection have genus $g'$.

If we let $\varepsilon_j$ denote the class of a fiber of the ruling on
$S_j$, it follows that
\eqn\fibrspn{
S_j\cdot\varepsilon_k=\cases{-2 & if $k=j$ \crcr 1 & if $|k-j|=1$,
$k\ne N$ \crcr 2 & if $k=j+1=N$ \crcr 0 & otherwise, \crcr} } which
reproduces the Cartan matrix for $Sp(N)$ (with the usual sign change
needed when passing between the conventions of Lie group theory and
algebraic geometry).  The configuration of surfaces is illustrated in
figure 1.  (Dotted lines in the figures are intended as an aid to
visualization, and have no particular meaning; thick lines indicate
the curves along which pairs of surfaces meet; thin lines indicate
curves which are contained in only one surface.)

\iffigs
\midinsert
\centerline{\epsfxsize=4.14in\epsfbox{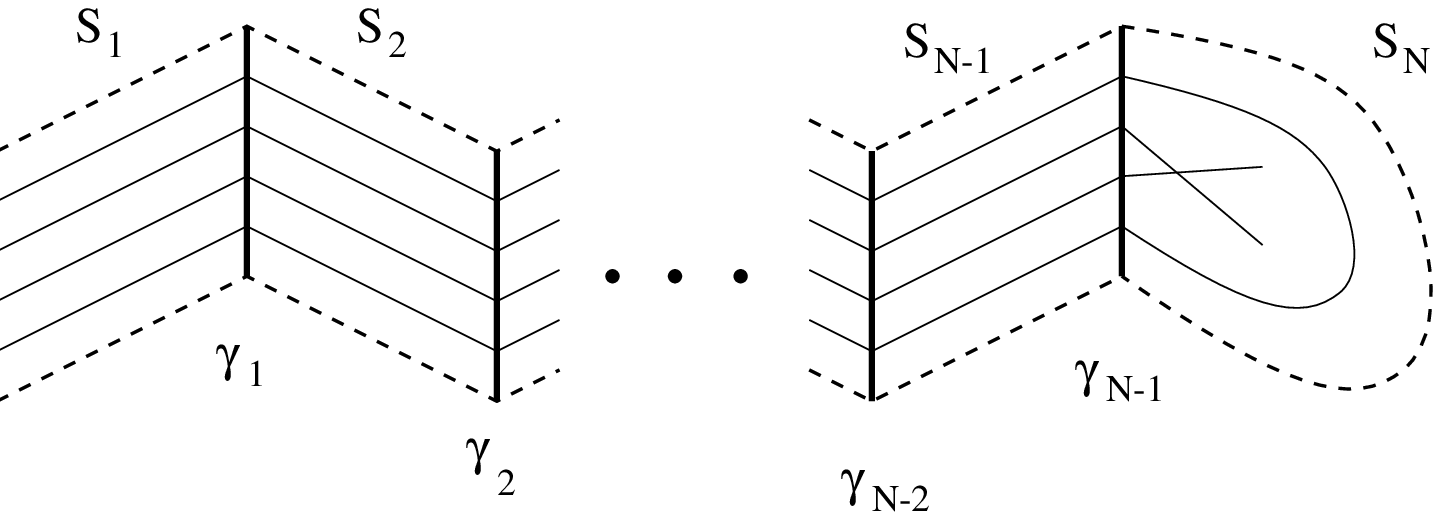}}
\centerline{Figure 1. The configuration of surfaces yielding $Sp(N)$ with
$n_F$ fundamentals.}
\endinsert
\fi

Following \rKMnew, we note that each of the adjoint weights (which can
all be written as combinations of the $\varepsilon_k$'s) is
responsible for either $g$ or $g'$ hypermultiplets of charged matter,
depending on whether the corresponding parameter curve has genus $g$
or $g'$.  These hypermultiplets can be collected into $g$ copies of
the adjoint representation and $g'-g$ copies of the antisymmetric
representation.  In fact, this is the only way that the antisymmetric
representation can arise, so we find that $n_A=g'-g$.

For each copy of the fundamental representation in the charged matter
spectrum (each one filling out a half-hypermultiplet), let
$\sigma^{(\alpha)}_N$ be the lowest weight in the representation, so
that the other weights are given by
\eqn\sigmaspn{
\sigma_k^{(\alpha)}=\varepsilon_k+
\varepsilon_{k+1}+\dots+\varepsilon_{N-1}+
\sigma_N^{(\alpha)}.
}
These classes intersect the surfaces $S_j$ according to
\eqn\rtwtspn{
S_j\cdot\sigma_k^{(\alpha)}=\cases{ -1 & if $k=j$ \crcr 1 & if
$k=j+1$ \crcr 0 & if $k\ne j$, $j+1$. \crcr}
}
(As a partial converse, we can write
\eqn\sigmaspnconverse{
\varepsilon_k=\sigma_k^{(\alpha)}-\sigma_{k+1}^{(\alpha)}
}
for $k\le N-1$.)

Thus, if we let $S:=\sum_{j=1}^{N} \varphi_j
S_j$ be an arbitrary divisor supported on the exceptional locus,
and introduce coordinates
$a_k=-S\cdot\sigma_k^{(\alpha)}=\varphi_k-\varphi_{k-1}$ on the space of
such divisors
(setting $\varphi_0=0$ for convenience), we can describe the negative of
the relative K\"ahler cone as being contained in another cone:
\eqn\relkkspn{
-\cK(X/\overline{X})\subseteq\{S=\sum\varphi_jS_j\ |\
-S\cdot\varepsilon_k>0 \} =\{S\ |\ a_1>a_2>\cdots>a_N>0\}, } where in
the last equality we used \sigmaspnconverse, the definition of $a_k$,
and the fact that $-S\cdot\varepsilon_N=2a_N$.  In other words,
$-\cK(X/\overline{X})$ is contained in the usual Weyl chamber for
$Sp(N)$.

The weights $\sigma^{(\alpha)}_k$ all define positive functions on the
Weyl chamber, so we will have $-S\cdot \sigma^{(\alpha)}_k>0$ whenever
$S\in-\cK(X/\overline{X})$.  This tells us that the negative of the
relative K\"ahler cone {\it is}\/ the Weyl chamber, and also that all
of the classes $\sigma^{(\alpha)}_k$ are represented by unions of
holomorphic curves.  In fact, $\sigma^{(\alpha)}_N$ will be
represented by an irreducible holomorphic curve, while
$\sigma^{(\alpha)}_k$ for $k<N$ will be represented by a reducible
curve as specified in \sigmaspn, with the fibers being chosen so that
the resulting curve is connected.  (Such fibers appear in figure 1 as
thin lines.)

Note that $\sigma^{(\alpha)}_N$ is an exceptional curve of the first
kind on $S_N$.  The fiber which contains it must be reducible, with
another component
$\sigma^{(\alpha')}_N:=\varepsilon_N-\sigma^{(\alpha)}_N$ which is
also an exceptional curve of the first kind, and which also serves as
the lowest weight for a fundamental half-hypermultiplet. (One such
pair is illustrated in figure 1.)  Thus, we see that {\it the
fundamental representations must occur in pairs}, i.e., the number of
half-hypermultiplets is even, confirming one of our predictions {}from
the field theory analysis.  There are $2n_F$ exceptional curves of the
first kind contained in fibers, combining to give $n_F$ reducible
fibers on the ruled surface $S_N$.

We can compute the cubic intersection form as follows.  First, on a
minimal ruled surface $S$ over a holomorphic curve of genus $g$, any
$2$-section $\gamma$ of genus $g'$ must satisfy $\gamma^2=4(g'-2g+1)$.
To apply this in our situation, we must modify the formula by the
$n_F$ blowups to which $S_N$ has been subjected; the result is
\eqn\newspn{
S_{N-1}^2S_N= (\gamma_{N-1})^2_{S_N}=4(g'-2g+1)-n_F=(8-8g-n_F)+(4g'-4).
}

Second, the holomorphic curve $\gamma_j=S_j\cap S_{j+1}$ has genus
$g'$, so its normal bundle must have degree $2g'-2$; this implies that
\eqn\curvespn{
S_j^2S_{j+1}+S_jS_{j+1}^2=(\gamma_j)^2_{S_{j+1}}+(\gamma_j)^2_{S_j}=2g'-2.
}

Third, since each $S_j$ for $j<N$ is a {\it minimal}\/ ruled surface
with disjoint sections we find
\eqn\surfspn{
 S_{j-1}^2S_j+S_jS_{j+1}^2=(\gamma_{j-1})^2_{S_j}+
(\gamma_j)^2_{S_j}=0 \qquad \hbox{for all $j\le N-1$}.
}

Now an easy induction based on \newspn, \curvespn, and \surfspn\
yields the formula
\eqn\aabspn{\eqalign{
S_j^2S_{j+1}&=(j-N+3)(2g'-2) + (8-8g-n_F) \cr
S_jS_{j+1}^2&=(N-j-2)(2g'-2) - (8-8g-n_F). \cr }} The only other
intersection numbers which are non-zero are determined by the fact
that the complex surfaces $S_j$ for $j<N$ are minimal ruled surfaces
over a holomorphic curve of genus $g'$, whereas $S_N$ is ruled over a
holomorphic curve of genus $g$ with $n_F$ reducible fibers.  This
implies that
\eqn\cubsispn{
S_j^3=(K_{S_j})^2_{S_j}=\cases{8-8g' & if $j<N$\crcr
8-8g-n_F & if $j=N$.\crcr}
}

We can assemble \cubsispn\  and \aabspn\ into a single formula for
the cubic form:
\eqn\cubicspn{\eqalign{
S^3={}&(\sum_{j=1}^{N}\varphi_jS_j)^3
\cr{}={}& (8-8g-n_F) \varphi_N^3+(8-8g')\sum_{j=1}^{N-1}\varphi_j^3
+ 3(g'-1)\sum_{j=1}^{N-1}(\varphi_j^2\varphi_{j+1}
                           +\varphi_j\varphi_{j+1}^2)
\cr&+3\sum_{j=1}^{N-1}((8-8g-n_F)+(2j+5-2N)(g'-1))
      (\varphi_j^2\varphi_{j+1}-\varphi_j\varphi_{j+1}^2).
}}

To put this into a more convenient form, we need two algebraic
identities, which can be established easily by induction {}from our
basic relations $a_j=\varphi_j-\varphi_{j-1}$, $\varphi_0=0$.
\eqna\identities
$$\eqalignno{
     \sum_{j=k+1}^Na_j^3={}& -\varphi_k^3+\varphi_N^3+
     3\sum_{j=k}^{N-1}(\varphi_j^2\varphi_{j+1}
                       -\varphi_j\varphi_{j+1}^2)&\identities a\cr
\sum_{1\le i<j\le N}\left((a_i-a_j)^3+(a_i+a_j)^3\right)
={}&8\sum_{j=1}^{N-1}\varphi_j^3
     -3\sum_{j=1}^{N-1}(\varphi_j^2\varphi_{j+1}+\varphi_j\varphi_{j+1}^2)
     &\identities b\cr&
     +3\sum_{j=1}^{N-1}(2N-2j-5)(\varphi_j^2\varphi_{j+1}
     -\varphi_j\varphi_{j+1}^2). }$$ Then \cubicspn\ can be rewritten
     as
\eqn\cubicspnbis{\eqalign{
S^3 =& (8-8g-n_F)\sum_{j=1}^N a_j^3 + (1-g')\sum_{1\le i<j\le
N}\left((a_i-a_j)^3+(a_i+a_j)^3\right) \cr =& (1-g)\left(8\sum_{j=1}^N
a_j^3 +\sum_{1\le i<j\le N}\left((a_i-a_j)^3+(a_i+a_j)^3\right)
\right)
\cr&
-n_F\left(\sum_{j=1}^N a_j^3 \right) -n_A\left(\sum_{1\le i<j\le
N}\left((a_i-a_j)^3+(a_i+a_j)^3\right)\right), }} where in the last
line we used the fact that $n_A=g'-g$.  This formula is in perfect
agreement with \spnprep, since the prepotential in the Calabi--Yau
theories is given by $\cF={1\over6}S^3$.

\subsec{$SU(N)$}

We next turn to the case of gauge group $SU(N)$, with hypermultiplets
in $g$ copies of the adjoint representation and $n_F$ copies of the
fundamental representation. (We omit the antisymmetric representation
in order to keep the geometry simple.)  Producing $SU(N)$ gauge
symmetry is relatively straightforward: we need $N-1$ complex surfaces
$S_j$ which shrink to zero size, with each $S_j$ being ruled over a
holomorphic curve of genus $g$.  The surfaces $S_j$ and $S_{j+1}$ must
meet along a holomorphic curve $\gamma_j$ which is a section for the
ruling on each of them.

If we let $\varepsilon_j$ denote the class of a fiber of the ruling on
$S_j$, it follows that
\eqn\fibrsun{
S_j\cdot\varepsilon_k=\cases{-2 & if $k=j$ \crcr 1 & if $|k-j|=1$
\crcr 0 & if $|k-j|>1$, \crcr} } which reproduces the negative of the
Cartan matrix for $SU(N)$.  The configuration of surfaces is
illustrated in figure 2.

\iffigs
\midinsert
\centerline{\epsfxsize=3.96in\epsfbox{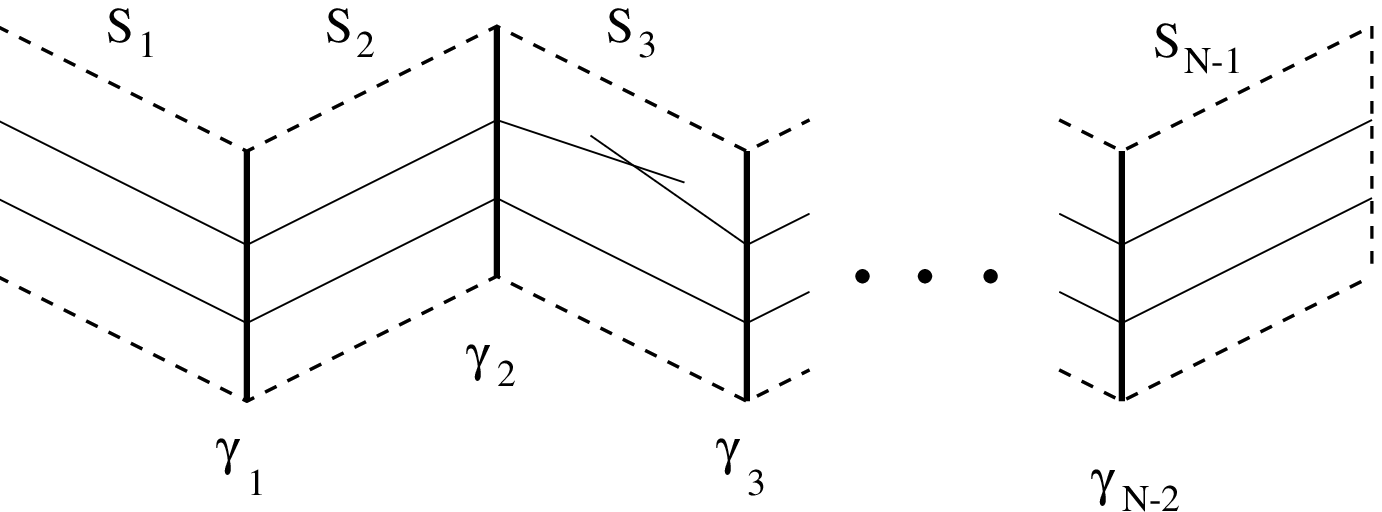}}
\centerline{Figure 2. The configuration of surfaces yielding $SU(N)$ with
$n_F$ fundamentals.}
\endinsert
\fi

For each copy of the fundamental representation in the charged matter
spectrum, let $\sigma^{(\alpha)}_N$ be the lowest weight in the
representation, so that the complete set of weights in the
representation are given by
\eqn\sigmasunbis{
\sigma_k^{(\alpha)}=\varepsilon_k+
\varepsilon_{k+1}+\dots+\varepsilon_{N-1}+
\sigma_N^{(\alpha)}.
} (Note that the weights occurring in the charged matter spectrum are
then $\pm
\sigma_k^{(\alpha)}$, since the spectrum contains both the
fundamental representation and its complex conjugate, pairing up to form
hypermultiplets.)
These classes intersect the surfaces $S_j$ according to
\eqn\rtwtsun{
S_j\cdot\sigma_k=\cases{ -1 & if $k=j$ \crcr 1 & if $k=j+1$ \crcr 0 &
if $k\ne j$, $j+1$. \crcr} }
(Conversely, we can write
\eqn\sigmasunconverse{
\varepsilon_k=\sigma_k^{(\alpha)}-\sigma_{k+1}^{(\alpha)}
}
for $k\le N-1$.)

Thus, if we let $S:=\sum_{j=1}^{N-1} \varphi_j S_j$ be an arbitrary
divisor supported on the exceptional locus, and introduce coordinates
$a_k=-S\cdot\sigma_k^{(\alpha)}=\varphi_k-\varphi_{k-1}$ on the space
of such divisors (setting $\varphi_0=\varphi_N=0$ for convenience), we
can describe the negative of the relative K\"ahler cone as being
contained in another cone:
\eqn\relkksun{
-\cK(X/\overline{X})\subseteq\{S=\sum\varphi_jS_j\ |\
-S\cdot\varepsilon_k>0 \} =\{S\ |\ a_1>a_2>\cdots>a_N\}, } where in
the last equality we used \sigmasunconverse\ and the definition of
$a_k$.  In other words, $-\cK(X/\overline{X})$ is contained in the
usual Weyl chamber for $SU(N)$.

At any given point in the Weyl chamber, the functions defined by the
weights $\sigma^{(\alpha)}_k$ are positive for small values of $k$, and
negative for large values of $k$.  So on the cone
$-\cK(X/\overline{X})$, there must be some index $\ell$ such that
$a_\ell>0>a_{\ell+1}$ on that cone.  That condition identifies
$-\cK(X/\overline{X})$ as one of the sub-wedges we encountered when
studying the gauge theory.

The classes represented by connected holomorphic curves on $X$ are
then $\sigma_1$, \dots, $\sigma_\ell$ and $-\sigma_{\ell+1}$, \dots,
$-\sigma_N$.  These can all be written in the form
\eqn\sigmasun{\eqalign{
\sigma^{(\alpha)}_k=\varepsilon_k+
\varepsilon_{k+1}+\dots+\varepsilon_{\ell-1}+
\sigma^{(\alpha)}_\ell &\qquad \hbox{for $k\le \ell$}\cr
-\sigma^{(\alpha)}_k= -\sigma^{(\alpha)}_{\ell+1}+
\varepsilon_{\ell+1}+
\varepsilon_{\ell+2}+ \dots+
\varepsilon_{k-1} &\qquad \hbox{for $k\ge \ell+1$},\cr
}} so we see that $\sigma^{(\alpha)}_\ell$ and
$-\sigma^{(\alpha)}_{\ell+1}$ are represented by irreducible
holomorphic curves.  Moreover, the remaining fiber class can be
written as
$\varepsilon_\ell=\sigma^{(\alpha)}_\ell+(-\sigma^{(\alpha)}_{\ell+1})$,
so we see that there are $n_F$ reducible fibers in the ruling on
$S_\ell$, while the other $S_j$'s are minimal ruled surfaces.  The
fibers and reducible fibers which account for the matter
representation are illustrated in figure 2 with thin lines.

The other sub-wedges within the K\"ahler cone must be related to the
given one by performing flops.  Explicitly in this case, if we flop
the curves $\sigma_\ell^{(\alpha)}$, we move the reducible fibers {}from
$S_\ell$ to $S_{\ell-1}$, whereas if we flop the curves
$-\sigma_{\ell+1}^{(\alpha)}$ we move the reducible fibers {}from
$S_\ell$ to $S_{\ell+1}$.

We can compute the cubic intersection form and complete the
specification of the geometric data as follows.  First, the
holomorphic curve $\gamma_j=S_j\cap S_{j+1}$ has genus $g$, so its
normal bundle must have degree $2g-2$; this implies that
\eqn\curvesun{
S_j^2S_{j+1}+S_jS_{j+1}^2=(\gamma_j)^2_{S_{j+1}}+(\gamma_j)^2_{S_j}=2g-2.
}

Second, since each $S_j$ for $j\ne\ell$ is a minimal ruled surface
with disjoint sections while $S_\ell$ has been blown up $n_F$ times
{}from a minimal ruled surface, we find
\eqn\surfsun{
 S_{j-1}^2S_j+S_jS_{j+1}^2=(\gamma_{j-1})^2_{S_j}+
(\gamma_j)^2_{S_j}=\cases{0 & if $j\ne \ell$ \crcr -n_F & if
$j=\ell$.\crcr} }

These two relations together determine these intersection numbers up
to one undetermined constant---there is no analogue of \newspn\
available to us in this case.  That is, although the relative values
of the self-intersections of the holomorphic curves $\gamma_j$ have
been determined, the actual values have not been.  To specify these,
we define
\eqn\constsun{
c':=\cases{ (N-2)(g-1) - S_1S_2^2 & if $\ell>1$ \crcr (N-2)(g-1) - n_F
      - S_1S_2^2 & if $\ell=1$ .\crcr }} Then an easy induction based
      on \curvesun, \surfsun, and \constsun\ shows that
\eqn\aabsun{\eqalign{
S_j^2S_{j+1}&=\cases{ (2j+2-N)(g-1) + c' & if $j<\ell$ \crcr
               (2j+2-N)(g-1) + n_F + c' & if $j\ge \ell$, \crcr}\cr
               S_jS_{j+1}^2&=\cases{ (N-2j)(g-1) - c' & if $j<\ell$
               \crcr (N-2j)(g-1) - n_F - c' & if $j\ge \ell$,
               \crcr}\cr }} so that all values are determined by this
               one choice.

Finally, since the complex surfaces $S_j$ are ruled over a holomorphic
curve of genus $g$, and all of them except for $S_\ell$ are minimal
ruled surfaces, whereas $S_\ell$ has $n_F$ reducible fibers, it
follows that
\eqn\cubsisun{
S_j^3=(K_{S_j})^2_{S_j}=\cases{8-8g & if $j\ne \ell$\crcr 8-8g-n_F &
if $j=\ell$.\crcr} } We can assemble \aabsun\ and \cubsisun\ into a
single formula for the cubic form:
\eqn\cubicsun{\eqalign{
S^3=(\sum_{j=1}^{N-1}\varphi_jS_j)^3 ={}& -n_F
\varphi_\ell^3+(8-8g)\sum_{j=1}^{N-1}\varphi_j^3 +
3(g-1)\sum_{j=1}^{N-2}(\varphi_j^2\varphi_{j+1}+\varphi_j\varphi_{j+1}^2)
\cr&+3\sum_{j=1}^{N-2}(c'+(2j+1-N)(g-1))
      (\varphi_j^2\varphi_{j+1}-\varphi_j\varphi_{j+1}^2)
\cr&+3n_F
     \sum_{j=\ell}^{N-2}(\varphi_j^2
\varphi_{j+1}-\varphi_j\varphi_{j+1}^2).
}} To convert this expression into the $a_j$ coordinates, we need one
additional algebraic identity, valid when $\varphi_N=0$:
\eqn\newident{\eqalign{
\sum_{1\le i<j\le N}(a_i-a_j)^3={}&8\sum_{j=1}^{N-1}\varphi_j^3
     -3\sum_{j=1}^{N-1}(\varphi_j^2\varphi_{j+1}+\varphi_j\varphi_{j+1}^2)
\cr&
+3\sum_{j=1}^{N-1}(N-2j-1)(\varphi_j^2\varphi_{j+1}
                            -\varphi_j\varphi_{j+1}^2). }} Using this
                            together with \identities{a,b}\ and
                            exploiting the fact that $\varphi_N=0$, we
                            can rewrite \cubicsun\ as
\eqn\cubicsunbis{
S^3 = -n_F \sum_{j=\ell+1}^N a_j^3 + (1-g) \sum_{1\le i<j\le
N}(a_i-a_j)^3 + c' \sum_{j=1}^N a_j^3 } or, using the fact that
$a_\ell>0>a_{\ell+1}$ and so $\sum_{j=\ell+1}^N a_j^3 =
-{1\over2}\left(\sum_{j=1}^N a_j^3 - \sum_{j=1}^N |a_j|^3 \right)$, as
\eqn\cubicsunter{
S^3= (1-g)\sum_{i<j}(a_i-a_j)^3 + (c'+{n_F\over2}) \sum_{j=1}^Na_j^3
 -{n_F\over2} \sum_{j=1}^N |a_j|^3.  } In this latter form, we have a
 single formula, valid throughout the Weyl chamber.

Comparing with \sunprep, we see that $c=c'+{n_F\over2}$ must be the
coefficient of the Chern--Simons term in the field theory.  (In fact,
it was the presence of the additional degree of freedom $c'$ in
specifying these $SU(N)$ Calabi--Yau theories which led us to discover
the possibility of such a Chern--Simons term.)  We immediately recover
the condition that $c+{n_F\over2}$ must be an integer.

\subsec{$Spin(2N+1)$}

We now consider the gauge group $Spin(2N+1)$, with hypermultiplets in
$g$ copies of the adjoint representation, and $n_V$ copies of the
vector representation.  (To keep the geometry relatively simple, we do
not allow matter in spinor representations, in either this case or the
next one.)  In order to produce $Spin(2N+1)$ gauge symmetry when the
ruled surfaces $S_j$ shrink to zero size,\foot{There is another way to
get $Spin(2N+1)$ gauge symmetry \rKMnew, but it leads to charged
matter in the antisymmetric rather than the vector representation.}
all but one of them must be ruled over a holomorphic curve of genus
$g$, while the last surface---labeled $S_N$ below---must be ruled over
a holomorphic curve of genus $g'$ which has a two-to-one map to the
curve of genus $g$.  (This is the opposite of what happened with
$Sp(N)$, corresponding to the exchange of long and short roots.)  The
surfaces $S_j$ and $S_{j+1}$ must meet along a holomorphic curve
$\gamma_j$.  When $j<N-1$ this curve is a section for the ruling on
each surface, while $\gamma_{N-1}$ is a $2$-section on $S_{N-1}$ and a
section on $S_N$.  The curve of intersection $\gamma_j$ has genus $g$
for $j<N-1$, while $\gamma_{N-1}$ has genus $g'$.

If we let $\varepsilon_j$ denote the class of a fiber of the ruling on
$S_j$, it follows that
\eqn\fibrsoodd{
S_j\cdot\varepsilon_k=\cases{-2 & if $k=j$ \crcr 1 & if $|k-j|=1$,
$j\ne N$ \crcr 2 & if $k+1=j=N$ \crcr 0 & otherwise. \crcr} } which
reproduces the negative of the Cartan matrix for $Spin(2N+1)$.  The
configuration of surfaces is illustrated in figure 3.

\iffigs
\midinsert
\centerline{\epsfxsize=4.83in\epsfbox{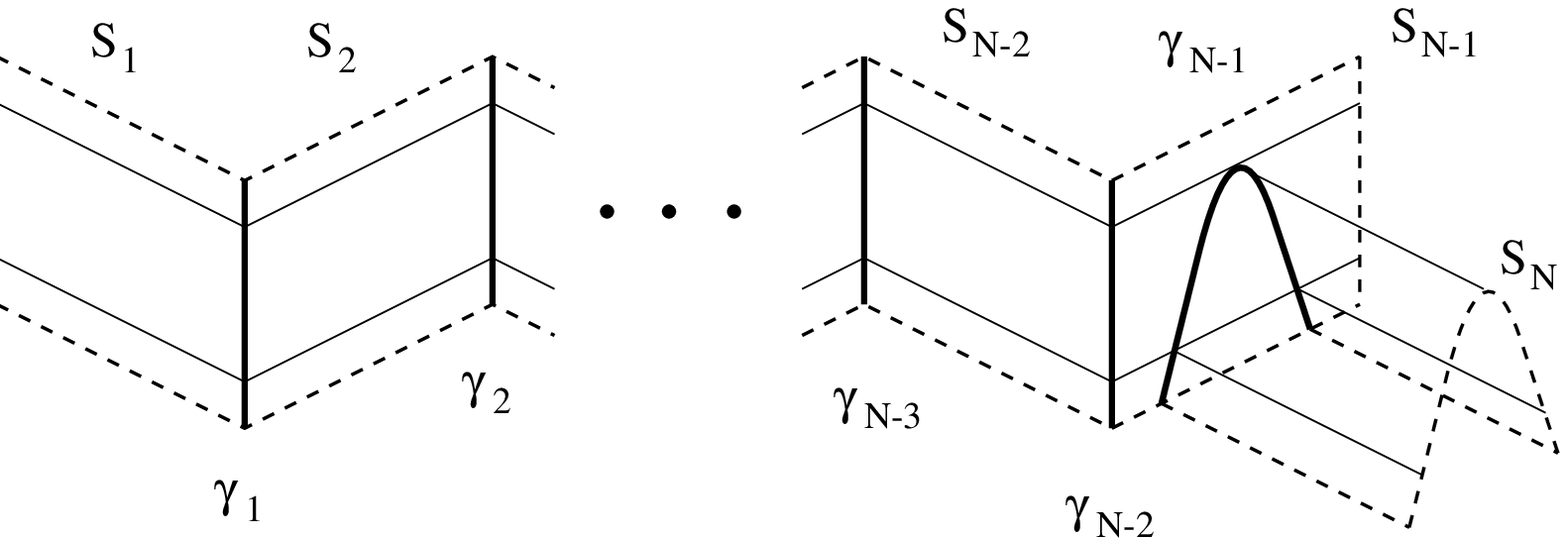}}
\centerline{Figure 3. The configuration of
surfaces yielding $Spin(2N+1)$ with
$n_V$ vectors.}
\endinsert
\fi

Following \rKMnew, we note that each of the adjoint weights (which can
all be written as combinations of the $\varepsilon_k$'s) is
responsible for either $g$ or $g'$ hypermultiplets of charged matter,
depending on whether the corresponding parameter curve has genus $g$
or $g'$.  These hypermultiplets can be collected into $g$ copies of
the adjoint representation and $g'-g$ copies of the vector
representation.  In fact, this is the only way that the vector
representation can arise, so we find that $n_V=g'-g$.  Since this is
the only matter content we have allowed, there will be no singular
fibers in any of the rulings.

If we define
\eqn\sigmasoodd{
\sigma_k = \varepsilon_k+\dots+\varepsilon_N,
} then the classes in the vector representation consist of $\pm
\sigma_1$, \dots, $\pm \sigma_N$ and a neutral class.  (These can be
realized geometrically if we choose the fiber $\varepsilon_{N-1}$ to
pass through one of the ramification points of the two-to-one map
$\gamma_{N-1}\to
\gamma_{N-2}$, as
illustrated in figure 3.)  Conversely, we can write
\eqn\sigmasooddconverse{
\varepsilon_k=\sigma_k-\sigma_{k+1}
}
for $k\le N$, setting $\varepsilon_{N+1}=0$.

For ease of computation, we introduce $\widetilde{S}_j$, defined as
\eqn\stildesoodd{
\widetilde{S}_j:=\cases{S_j & if $j<N$ \crcr {1\over2}S_N & if $j=N$.}
} The intersection properties of these classes are
\eqn\rtwtsooddbis{
\widetilde{S}_j\cdot\sigma_k=\cases{ -1 & if $k=j$ \crcr 1 & if
$k=j+1$ \crcr 0 & if $k\ne j$, $j+1$. \crcr}
}
Thus, if we let let $S:=\sum_{j=1}^{N} \varphi_j
\widetilde{S}_j$ be an arbitrary divisor supported on the exceptional
locus, and introduce coordinates
$a_k=-S\cdot\sigma_k=\varphi_k-\varphi_{k-1}$ on the space of such
divisors (setting $\varphi_0=0$ for convenience), we can describe the
negative of the relative K\"ahler cone as being contained in another
cone:
\eqn\relkksoodd{
-\cK(X/\overline{X})
\subseteq\{S=\sum\varphi_jS_j\ |\ -S\cdot\varepsilon_k>0 \}
=\{S\ |\ a_1>a_2>\cdots>a_N>0\}, } where in the last equality we used
\sigmasooddconverse\ and the definition of $a_k$.  In other words,
$-\cK(X/\overline{X})$ is contained in the usual Weyl chamber for
$Spin(2N+1)$.  In fact, since we have no singular fibers, we have
accounted for all of the conditions and these two cones must coincide.

We can compute the cubic intersection form as follows.  First, since
$S_{N-1}$ is a minimal ruled surface over a holomorphic curve of genus
$g$, and $\gamma_{N-1}$ is a $2$-section on $S_{N-1}$, we have
\eqn\relA{
S_{N-1}S_{N}^2 = (\gamma_{N-1})^2_{S_{N-1}} = 4(g'-2g+1) =
(4g'-4)-(8g-8).  } Moreover, since $\gamma_{N-1}$ has genus $g'$, its
normal bundle has degree $2g'-2$, {}from which it follows that
\eqn\curvesooddbis{
S_{N-1}^2S_{N}+S_{N-1}S_{N}^2
=(\gamma_{N-1})^2_{S_{N}}+(\gamma_{N-1})^2_{S_{N-1}}=2g'-2.
}
Putting these together, we infer
\eqn\relB{
S_{N-1}^2S_N = (8g-8)-(2g'-2).  } Calculating in terms of the
$\widetilde{S}_j$'s, we find
\eqn\relD{
\widetilde{S}_{N-1}^2\widetilde{S}_N = {1\over2}S_{N-1}^2S_{N} =
(4g-4)+(1-g') } and
\eqn\relC{
\widetilde{S}_{N-1}^2\widetilde{S}_{N}+
\widetilde{S}_{N-1}\widetilde{S}_{N}^2
= {1\over2}S_{N-1}^2S_{N}+{1\over4}S_{N-1}S_{N}^2 = 2g-2.
}

Second, the holomorphic curve $\gamma_j=S_j\cap S_{j+1}$ has genus $g$
for $j\le N-2$, so its normal bundle must have degree $2g-2$; this
implies that
\eqn\curvesoodd{
S_j^2S_{j+1}+S_jS_{j+1}^2=(\gamma_j)^2_{S_{j+1}}+(\gamma_j)^2_{S_j}=2g-2,
\quad \hbox{for $j\le N-2$}.
} Combined with \relC, we find
\eqn\curvesooddter{
\widetilde{S}_j^2\widetilde{S}_{j+1}+
\widetilde{S}_j\widetilde{S}_{j+1}^2=2g-2,
\quad \hbox{for $j\le N-1$}.
}

Third, the ruled surface $S_{N-1}$ has a $2$-section $\gamma_{N-1}$
and a section $\gamma_{N-2}$ which are disjoint.  Writing the
$2$-section in the form
$\gamma_{N-1}=2\gamma_{N-2}+k\varepsilon_{N-1}$ (in $H^2(S_{N-1})$),
we find the relation $2\gamma_{N-2}^2+k=0$, so that
$\gamma_{N-1}^2=4\gamma_{N-2}^2+4k=-4\gamma_{N-2}^2$.  In other words,
\eqn\relone{
4S_{N-2}^2S_{N-1}+S_{N-1}S_N^2=
4(\gamma_{N-2})^2_{S_{N-1}}+(\gamma_{N-1})^2_{S_{N-1}}=0,
}
{}from which it follows that
\eqn\relonebis{
\widetilde{S}_{N-2}^2\widetilde{S}_{N-1}+
\widetilde{S}_{N-1}\widetilde{S}_N^2
=0.  } In addition, since $S_j$ is a minimal ruled surface with
disjoint sections for $j\le N-2$, we find
\eqn\surfsoodd{
 S_{j-1}^2S_j+S_jS_{j+1}^2=(\gamma_{j-1})^2_{S_j}+
(\gamma_j)^2_{S_j}=0, \quad \hbox{for $j\le N-2$}.
}
Combined with \relonebis, this implies
\eqn\surfsooddbis{
\widetilde{S}_{j-1}^2\widetilde{S}_j+
\widetilde{S}_j\widetilde{S}_{j+1}^2=0,
 \quad \hbox{for $j\le N-1$}.  }

Finally, all of the $S_j$'s being minimal ruled surfaces implies that
\eqn\cubsisoodd{
S_j^3=(K_{S_j})^2_{S_j}=\cases{8-8g & if $j<N$\crcr 8-8g' & if
$j=N$,\crcr} } or in other words,
\eqn\cubsisooddbis{
\widetilde{S}_j^3=\cases{8-8g & if $j<N$\crcr
1-g' & if $j=N$.\crcr}
}

The relations
\relD, \curvesooddter, \surfsooddbis, and \cubsisooddbis\ are precisely
parallel to the relations
\newspn, \curvespn, \surfspn, and \cubsispn,
with the substitution of $1-g'$ for $8-8g-n_F$ and $g$ for $g'$.
We can thus immediately apply the results of our computation {}from the
$Sp(N)$ case, and conclude that
\eqn\aabsoodd{\eqalign{
\widetilde{S}_j^2\widetilde{S}_{j+1}&=(j-N+3)(2g-2) + (1-g')  \cr
\widetilde{S}_j\widetilde{S}_{j+1}^2&=(N-j-2)(2g-2) - (1-g'), \cr
}}
as well as the final result
\eqn\cubicsoodd{\eqalign{
S^3 =& (1-g')\sum_{j=1}^N a_j^3 + (1-g)\sum_{1\le i<j\le
N}\left((a_i-a_j)^3+(a_i+a_j)^3\right) \cr
=& (1-g)\left(\sum_{j=1}^N a_j^3 +\sum_{1\le i<j\le
N}\left((a_i-a_j)^3+(a_i+a_j)^3\right) \right)
\cr&
-n_V\left(\sum_{j=1}^N a_j^3 \right) }} where in the last line we used
the fact that $n_V=g'-g$.  Since the prepotential is given by
$\cF={1\over6}S^3$, we again find perfect agreement with the field
theory formula \sonoprep.

\subsec{$Spin(2N)$}

Finally, we consider the gauge group $Spin(2N)$, with hypermultiplets
in $g$ copies of the adjoint representation, and $n_V$ copies of the
vector representation.  As in the $SU(N)$ case, producing $Spin(2N)$
gauge symmetry is relatively straightforward: all of the complex
surfaces $S_j$ must be ruled over a holomorphic curve of genus $g$,
with $S_j$ and $S_{j+1}$ meeting along a holomorphic curve $\gamma_j$
which is a section of the rulings on both ($j\le N-2$), and similarly
for $S_{N-2}$ and $S_N$ meeting along $\gamma_{N-1}$.

If we let $\varepsilon_j$ denote the class of a fiber of the ruling on
$S_j$, it follows that
\eqn\fibrsoeven{
S_j\cdot\varepsilon_k=\cases{-2 & if $k=j$ \crcr 1 & if $|k-j|=1$,
$j\ne N$, $k\ne N$ \crcr 1 & if $(j,k)=(N-2,N)$ or $(N,N-2)$ \crcr 0 &
otherwise. \crcr} } which reproduces the negative of the Cartan matrix
for $Spin(2N)$.

For each copy of the vector representation in the matter spectrum,
there are weights $\sigma_1^{(\alpha)}$, \dots, $\sigma_N^{(\alpha)}$,
related to the adjoint weights by
\eqn\sigmasoeven{ \sigma_k^{(\alpha)} =
\varepsilon_k+\cdots+\varepsilon_{N-1}+\sigma_N^{(\alpha)}, }
and
\eqn\sigmasoevenbis{
\sigma_{N-1}^{(\alpha)}+\sigma_N^{(\alpha)}=\varepsilon_N, }
such that $\pm \sigma_1^{(\alpha)}$, \dots, $\pm \sigma_N^{(\alpha)}$
are the weights occurring in the representation.  Conversely, the
$\varepsilon_k$'s can be determined {}from the $\sigma_k^{(\alpha)}$'s
by \sigmasoevenbis\ and
\eqn\sigmasoevenconverse{
\varepsilon_k=\sigma_k^{(\alpha)}-\sigma_{k+1}^{(\alpha)}
}
for $k\le N-1$.

Either $\sigma_N^{(\alpha)}$ or $-\sigma_N^{(\alpha)}$ must have
positive intersection number with classes $S\in -\cK(X/\overline{X})$.
In fact, since
$\sigma_N^{(\alpha)}={1\over2}(\varepsilon_N-\varepsilon_{N-1})$, by
exchanging labels on $S_{N-1}$ and $S_N$ if necessary, we can assume
that
\eqn\assump{
-S\cdot\sigma_N^{(\alpha)}=
-{1\over2}S\cdot(\varepsilon_N-\varepsilon_{N-1})>0 \hbox{ for } S\in
-\cK(X/\overline{X}).  } (Such an exchange of labels is permissible
due to the symmetry of the Dynkin diagram for $Spin(2N)$.)

Since the classes $\sigma_k^{(\alpha)}$ are all represented by
connected holomorphic curves---and according to \sigmasoeven\ all can
be so represented in terms of $\sigma_N^{(\alpha)}$ and various fibers
of rulings---we see that $\sigma_N^{(\alpha)}$ should actually be an
irreducible holomorphic curve.  Since $S_N\cdot
\sigma_N^{(\alpha)}=-1$, this curve must be contained in $S_N$ and be
an exceptional curve of the first kind there.  Moreover, since
$S_{N-1}\cdot \sigma_N^{(\alpha)}=1$ we see that $S_N$ and $S_{N-1}$
must meet each other.

{}From \sigmasoeven\ and \sigmasoevenbis\ we conclude that
$$ \varepsilon_N=\sigma_{N-1}^{(\alpha)} + \sigma_N^{(\alpha)}
=\varepsilon_{N-1} + 2\sigma_N^{(\alpha)}.$$ Thus, we see that there
is a particular fiber in the ruling on $\varepsilon_N$ which can be
written as $\delta^{(\alpha)}+2\sigma_N^{(\alpha)}$ for some
irreducible holomorphic curve $\delta^{(\alpha)}$, and
$\delta^{(\alpha)}$ also serves as a fiber in the ruling on $S_{N-1}$.
It is the curves $\delta^{(\alpha)}$ along which $S_N$ and $S_{N-1}$
meet.  This is illustrated in figure 4, where we show the
configuration of surfaces, explicitly showing only one curve
$\delta^{(\alpha)}$ of intersection.  The curve $\sigma_N^{(\alpha)}$
(which lies on $S_N$) is indicated with a thin line.

\iffigs
\midinsert
\centerline{\epsfxsize=4.53in\epsfbox{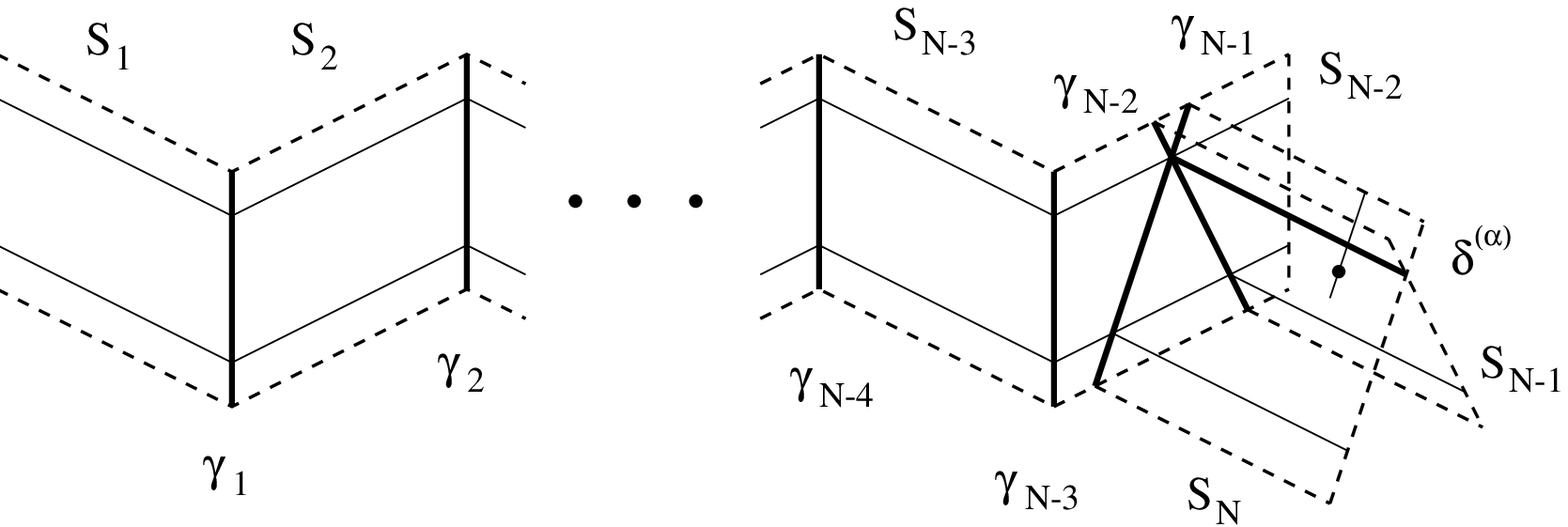}}
\centerline{Figure 4. The
configuration of surfaces yielding $Spin(2N)$ with
$n_V$ vectors.}
\endinsert
\fi

There is one additional detail about the configuration which we need.
Since $S_N\cdot \delta^{(\alpha)}=S_N\cdot \varepsilon_{N-1}=0$, we
see that $\delta^{(\alpha)}$ is a holomorphic curve whose
self-intersection on $S_N$ is $-2$. Under those circumstances, the
only way that $\delta^{(\alpha)}+2\sigma_N^{(\alpha)}$ can be a fiber
in the ruling is if the exceptional curve $\sigma_N^{(\alpha)}$ passes
through an ordinary double point, and indeed a local computation shows
that this happens
\rGMnew.  This implies that the exceptional curve $\sigma_N^{(\alpha)}$
does not arise {}from an ordinary blowup, but rather {}from one of the
``directed blowups of weight $1$'' introduced in \rBirGeoRDP.  The
singular point is indicated in the figure with a dot.

For ease of computation, we introduce $\widetilde{S}_j$, defined as
\eqn\stildesoeven{
\widetilde{S}_j:=\cases{S_j & if $j<N$ \crcr {1\over2}(S_N-S_{N-1}) & if
$j=N$.}  } The intersection properties of the classes
$\sigma_k^{(\alpha)}$ are
\eqn\rtwtsoevenbis{
\widetilde{S}_j\cdot\sigma_k^{(\alpha)}=\cases{ -1 & if $k=j$ \crcr 1 & if
$k=j+1$ \crcr 0 & if $k\ne j$, $j+1$. \crcr} } Thus, if we let
$S:=\sum_{j=1}^{N} \varphi_j
\widetilde{S}_j$ be an arbitrary divisor supported on the exceptional
locus, and introduce coordinates
$a_k=-S\cdot\sigma_k^{(\alpha)}=\varphi_k-\varphi_{k-1}$ on the space
of such divisors (setting $\varphi_0=0$ for convenience), we can
describe the negative of the relative K\"ahler cone as being contained
in another cone:
\eqn\relkksoeven{
-\cK(X/\overline{X})\subseteq\{S=\sum\varphi_jS_j\ |\
-S\cdot\varepsilon_k>0, -S\cdot \sigma_N^{(\alpha)}>0 \} =\{S\ |\
a_1>a_2>\cdots>a_N>0\}, } where in the last equality we used
\sigmasoevenconverse, \assump, and the definition of $a_k$.  The cone
on the right in \relkksoeven\ is the usual Weyl chamber for
$Spin(2N)$, and all of the weights $\sigma^{(\alpha)}_k$ define
positive functions on it; it follows that the two cones in
\relkksoeven\ are equal.  Note that in this case, in order to obtain
the Weyl chamber we had to use more conditions than simply
$-S\cdot\varepsilon_k>0$.

We can compute the cubic intersection form as follows.  First, all of
the $S_j$'s are ruled surfaces over a holomorphic curve of genus $g$.
These ruled surfaces are minimal if $j<N$, whereas $S_N$ is obtained
{}from a minimal ruled surface by making $n_V$ directed blowups of
weight $1$.  Since each of these directed blowups decreases the value
of $K_S^2$ by $2$, we conclude that
\eqn\cubsisoeven{
S_j^3=(K_{S_j})^2_{S_j}=\cases{8-8g & if $j<N$\crcr
8-8g-2n_V & if $j=N$.\crcr}
}

Second, since there are $n_V$ of the holomorphic curves
$\delta^{(\alpha)}$ which constitute the intersection of $S_{N-1}$ and
$S_N$, the numerical properties
\eqn\deltaprops{
S_{N-1}\cdot\delta^{(\alpha)}=-2, \qquad S_N\cdot\delta^{(\alpha)}=0 }
immediately imply
\eqn\propA{
S_{N-1}^2S_N=-2n_V, \qquad S_{N-1}S_N^2=0.
}
Among the conclusions we can draw {}from  \cubsisoeven\ and \propA\ are
\eqn\propB{
\widetilde{S}_{N-1}^2\widetilde{S}_N={1\over2}(S_{N-1}^2S_N-S_{N-1}^3)
=4g-4-n_V , }
\eqn\propC{
\widetilde{S}_{N-1}\widetilde{S}_N^2
={1\over4}(S_{N-1}S_N^2-2S_{N-1}^2S_N+S_{N-1}^3) =2-2g+n_V , } and
\eqn\cubsisoeventer{
\widetilde{S}_N^3={1\over8}(S_N^3-3S_{N-1}S_N^2+3S_{N-1}^2S_N-S_{N-1}^3)
=-n_V.
}
The last one implies
\eqn\cubsisoevenbis{
\widetilde{S}_j^3=\cases{8-8g & if $j<N$\crcr
-n_V & if $j=N$.\crcr} }

Third, the holomorphic curve $\gamma_j=S_j\cap S_{j+1}$ has genus $g$
for $j\le N-2$, so its normal bundle must have degree $2g-2$; this
implies that
\eqn\curvesoeven{
S_j^2S_{j+1}+S_jS_{j+1}^2=(\gamma_j)^2_{S_{j+1}}+(\gamma_j)^2_{S_j}=2g-2,
\quad \hbox{for $j\le N-2$}.
} Similarly, $\gamma_{N-1}=S_{N-2}\cap S_N$ has genus $g$, so
\eqn\curvesoevenbis{
S_{N-2}S_N^2+S_{N-2}^2S_N=2g-2.
}
Adding \propB\ and \propC\ also yields $2g-2$, so we find
\eqn\curvesoeventer{
\widetilde{S}_j^2
\widetilde{S}_{j+1}+\widetilde{S}_j\widetilde{S}_{j+1}^2=2g-2,
\quad \hbox{for $j\le N-1$}.
}

Finally, the minimal ruled surface $S_{N-2}$ has three sections
$\gamma_{N-3}$, $\gamma_{N-2}$, and  $\gamma_{N-1}$, which satisfy
\eqn\propD{
(\gamma_{N-2}\cdot\gamma_{N-1})_{S_{N-2}}=n_V, \qquad
(\gamma_{N-3}\cdot\gamma_{N-2})_{S_{N-2}}=
(\gamma_{N-3}\cdot\gamma_{N-1})_{S_{N-2}}=0 .  } {}From these, it
easily follows that $(\gamma_{N-3})_{S_{N-2}}^2=-n_V$,
$(\gamma_{N-2})_{S_{N-2}}^2=(\gamma_{N-1})_{S_{N-2}}^2=n_V$, or in
other words,
\eqn\propE{
S_{N-3}^2S_{N-2}=-n_V; \qquad
S_{N-2}S_{N-1}^2=S_{N-2}S_N^2=n_V.
}
Combining this with \propB\ and \curvesoeven\ we find
\eqn\propF{
\widetilde{S}_{N-2}^2\widetilde{S}_{N-1}+
\widetilde{S}_{N-1}\widetilde{S}_N^2
=S_{N-2}^2S_{N-1}+2-2g+n_V =-S_{N-2}S_{N-1}^2+n_V=0.  } On the other
hand, since $S_j$ is a minimal ruled surface with disjoint sections
$\gamma_{j-1}$ and $\gamma_j$ for $j\le N-2$, we find
\eqn\surfsoeven{
 S_{j-1}^2S_j+S_jS_{j+1}^2=(\gamma_{j-1})^2_{S_j}+
(\gamma_j)^2_{S_j}=0, \quad \hbox{for $j\le N-2$}.
}
Combined with \propF, this implies
\eqn\surfsoevenbis{
\widetilde{S}_{j-1}^2\widetilde{S}_j+\widetilde{S}_j
\widetilde{S}_{j+1}^2=0,
 \quad \hbox{for $j\le N-1$}.
}

Before exploiting the parallel with the $Sp(N)$ computation, we must
check that there are no other non-zero intersection numbers among the
$\widetilde{S}_j$'s.  This is less obvious than in previous cases,
since the $S_j$'s {\it do}\/ have some other non-zero intersection
numbers.  We already saw non-zero values of $S_{N-2}S_N^2$ and
$S_{N-2}^2S_N$ occurring in
\curvesoevenbis\
and \propE; the only other such non-zero
intersection number is
\eqn\triple{
S_{N-2}\cdot S_{N-1}\cdot S_N = n_V.  } To verify that the
corresponding intersection numbers of the $\widetilde{S}_j$'s vanish,
we compute using the definition of $\widetilde{S}_j$. On the one hand,
{}from \curvesoeven, \curvesoevenbis, and
\propE\
we find
\eqn\calcA{
\widetilde{S}_{N-2}^2\widetilde{S}_N = {1\over2}(S_{N-2}^2S_N -
S_{N-2}^2S_{N-1}) = {1\over2}((2g-2-S_{N-2}S_N^2) -
(2g-2-S_{N-2}S_{N-1}^2) = 0.
}
On the other hand, {}from \propE\ and \triple\ we find
\eqn\calcB{
\widetilde{S}_{N-2}\widetilde{S}_N^2 =
{1\over4}(S_{N-2}S_N^2 - 2S_{N-2}S_{N-1}S_N + S_{N-2}S_{N-1}^2) =
{1\over4}(n_V -2n_V + n_V)=0,
}
and also
\eqn\calcC{
\widetilde{S}_{N-2}\cdot\widetilde{S}_{N-1}\cdot\widetilde{S}_N
= {1\over2}(S_{N-2}S_{N-1}S_N - S_{N-2}S_{N-1}^2)
={1\over2}(n_V-n_V)=0.  }

Having checked that all other relations vanish, we now note that the
relations
\propB, \curvesoeventer, \surfsoevenbis,
and \cubsisoevenbis\ are precisely parallel to the relations
\newspn, \curvespn, \surfspn, and \cubsispn,
with the substitution of $-n_V$ for $8-8g-n_F$ and $g$ for $g'$.
We can thus again apply the results of our computation {}from the
$Sp(N)$ case, and conclude that
\eqn\aabsoeven{\eqalign{
\widetilde{S}_j^2\widetilde{S}_{j+1}&=(j-N+3)(2g-2) - n_V  \cr
\widetilde{S}_j\widetilde{S}_{j+1}^2&=(N-j-2)(2g-2) + n_V, \cr
}} as well as the final result
\eqn\cubicsoeven{\eqalign{
S^3 = (1-g)\left(\sum_{1\le i<j\le
N}\left((a_i-a_j)^3+(a_i+a_j)^3\right) \right) -n_V\left(\sum_{j=1}^N
a_j^3 \right) .}}  Once again we find perfect agreement with the field
theory formula \soneprep.

\newsec{Examples of the New Fixed Points {}from Calabi--Yau
Degenerations}

Having verified that our detailed descriptions of the geometry match
the gauge theory in every detail, it remains to consider the question
of when a strong coupling limit will exist.

The configurations of surfaces $S_j$ which we have considered have the
property that they can be contracted to a curve $\overline{C}$ of
singularities on a Calabi--Yau space $\overline{X}$.  The area of that
curve (calculated with respect to some volume-one K\"ahler metric)
corresponds to the classical gauge coupling $1/g_0^2$ \msdelp, so to
find a strong coupling limit we need to be able to shrink the curve
$\overline{C}$ to zero size.  Expressed a bit more directly, we need
to find a Calabi--Yau space $\widehat{X}$ with a map $X\to\widehat{X}$
under which the surfaces $S_j$ shrink to a single singular point
$\widehat{P}$.  In the case that there are several $X_\alpha$'s
resolving $\overline{X}$ (differing by flops), all of them must map to
$\widehat{X}$, giving different resolutions of the singular point
$\widehat{P}$.

We have seen in our field theory analysis that the convexity of the
prepotential $\cF$ is a necessary condition for a strong coupling
limit to exist.  That can be seen directly in the geometry as well.
For if we take any $S=\sum \psi_jS_j\in -\cK(X/\overline{X})$ with the
coefficients $\psi_j$ being rational numbers, then for a sufficiently
large positive integer $m$ there will be a nonsingular surface $L$ on
$X$ which is an ample divisor, such that $L$ and $-mS$ have the same
intersection numbers with all curves $\sigma$ contracted by
$\pi:X\to\overline{X}$.  If the surfaces $S_j$ can be contracted to a
point by some map $X\to\widehat{X}$, then the curves $C_j=L\cap S_j$
can be contracted to a point by a map $L\to\widehat{L}$ (where
$\widehat{L}$ is the image of $L$ in $\widehat{X}$).  However, a
necessary \mumford\ and sufficient \grauert\ condition for such a
contraction of $L$ to exist is that the intersection matrix $(C_i\cdot
C_j)$ should be negative definite.  Computing these intersection
numbers on $X$, we see that $(L\cdot S_i\cdot S_j)=(-mS\cdot S_i\cdot
S_j)$ should be negative definite, or in other words, that
$({\partial^2\cF\over\partial\psi_i\partial\psi_j}|_S)=(S\cdot
S_i\cdot S_j)$ should be positive definite.  If
$({\partial^2\cF\over\partial\psi_i\partial\psi_j}|_S)$ is positive
definite for all divisors $S=\sum\psi_jS_j$ in the cone for which the
$\psi_j$'s are rational, it must be positive definite throughout the
cone.  But this was one of the characterizations of convexity of the
prepotential.

A sufficient condition for the existence of such a contraction mapping
is also known in algebraic geometry \refs{\mori,\kawcone}: the classes
in $\cK(X/\overline{X})$ must restrict to positive classes (sometimes
called ``ample $\IR$-divisors'') on each of the surfaces $S_j$.
However, if $S_j$ is the blowup of $\IP^2$ at more than $8$ points (as
happens in some of our configurations), it contains an infinite number
of exceptional curves of the first kind.  This fact makes the
``sufficient condition'' quite difficult to check, in general, so we
shall have to be content with verifying it in a few examples.  The
field theory analysis, by renormalization group flow upon adding mass
terms for the matter, implies that the condition will also be
satisfied for examples with less matter.  These examples are enough to
conclude that most of the strong-coupling limit points which we have
predicted based on the convexity analysis do in fact occur.

Our strategy for verifying the ampleness of these classes is as
follows: we will analyze the divisors whose classes lie along the
edges of the relative K\"ahler cone $\cK(X/\overline{X})$, and show
that each of them contracts most of the $S_j$'s to curves or to
points, while serving as an ample divisor on some birational model of
each $S_j$.  {}From this analysis, it is easy to conclude that the
classes in the interior of the cone correspond to ample divisors.

\subsec{$Sp(N)$ with $n_A=1$}

In our first example, we consider $Sp(N)$ with $g=0$ adjoints, $n_A=1$
antisymmetric tensors, and $n_F\le7$ fundamentals.  The geometry is
represented by Figure 1, with $S_1$, \dots, $S_{N-1}$ being elliptic
ruled surfaces meeting along elliptic curves $\gamma_j$, and $S_N$
being a del Pezzo surface of degree $8-n_F$.  {}From \aabspn, we see
that
\eqn\sispna{
(\gamma_j^2)_{S_j}=S_jS_{j+1}^2=-(8-n_F), } for $j<N$.  Since $n_F<8$,
we can contract $\gamma_j$ on $S_j$ to obtain an elliptic cone
$\overline{S}_j$ for $j<N$.

The divisor $L_k:=\sum_{j=1}^{k-1}jS_j+\sum_{j=k}^NkS_j$ lies on that
edge of the cone $-\cK(X/\overline{X})$ for which
$a_1=\dots=a_k>a_{k+1}=\dots=a_N=0$.  We let $\gamma_0$ be a section of the
ruling on $S_1$ disjoint from $\gamma_1$, and note that $S_j|_{S_j}=K_{S_j}=
-\gamma_{j-1}-\gamma_j$ for $j<N$ and $S_N|_{S_N}=K_{S_N}=-\gamma_{N-1}$.
The restriction $-L_k|_{S_j}$ can be computed as follows:
\eqn\lksjspna{
-L_k|_{S_j}=\cases{\gamma_{j-1}-\gamma_j & $k>j$\crcr
\gamma_{j-1} & $k=j$ \crcr
0 & $k<j$. \crcr} } It is then easy to see that $-L_k|_{S_j}$ is a nef
divisor which contracts $S_j$ along its ruling (mapping to a curve)
for $j<k\le N$, maps $S_j$ to a point for $j>k$, is ample on the
elliptic cone $\overline{S}_k$ when $j=k<N$.  The remaining case,
$-L_N|_{S_N}$, is an anticanonical divisor on the del Pezzo surface
$S_N$; this is ample as well.

Generalizing the codimension $N=1$ result of \msdelp, there is a
string theory argument that the deformations of the singularity in
these cases, corresponding to the Higgs branch of the gauge theory in
the strong coupling limit, must be the same as the moduli space of
$N$, $E_{n_F+1}$ instantons.

\subsec{$Sp(N)$ with $n_A=0$}

In our next example, we consider $Sp(N)$ with $g=0$ adjoints, $n_A=0$
antisymmetric tensors, and $n_F=2N+4$ fundamentals.  The geometry is
again represented by Figure 1, but this time $S_1$, \dots, $S_{N-1}$
are minimal rational ruled surfaces meeting along rational curves
$\gamma_j$, and $S_N$ is the blowup of a minimal rational ruled
surface at $2N+4$ points lying on a $2$-section of the ruling.  {}From
\aabspn, we see that
\eqn\sispn{
(\gamma_{j-1}^2)_{S_j}=S_{j-1}^2S_{j}=-2j, } for $1<j\le N$.  We can
thus contract $\gamma_{j-1}$ on $S_j$ to obtain a rational cone
$\overline{S}_j$ for $1<j<N$.  We can also contract $\gamma_{N-1}$
(the proper transform of the $2$-section) on $S_N$ to obtain a surface
$\overline{S}_N$ which is somewhat more complicated than a cone.  Note
that $S_1$ can be either of the Hirzebruch surfaces $\IF_0$ or
$\IF_2$; we let $\overline{S}_1$ denote either $\IF_0$ or the
contraction of the negative section on $\IF_2$---a quadric in either
case.

The divisor $L_k:=\sum_{j=1}^{k-1}jS_j+\sum_{j=k}^NkS_j$ lies on that
edge of the cone $-\cK(X/\overline{X})$ for which
$a_1=\dots=a_k>a_{k+1}=\dots=a_N=0$.  We define
$\gamma_0=\gamma_1-2\varepsilon_1$ as a class on $S_1$, and note that
$S_j\cong\IF_{2j}$ for
$j>1$, and hence that $\gamma_{j-1} +2j\varepsilon_j \sim \gamma_j$ on
$S_j$.  The normal bundles are $S_j|_{S_j}=K_{S_j}=
-\gamma_{j-1}-\gamma_j-2\varepsilon_j$ for $j<N$ and
$S_N|_{S_N}=K_{S_N}=-\gamma_{N-1}+\varepsilon_N$, and the restriction
$-L_k|_{S_j}$ can be computed as follows:
\eqn\lksjspn{
-L_k|_{S_j}=\cases{0 & $k>j$\crcr
\gamma_{j} & $k=j<N$ \crcr
2k\varepsilon_j & $k<j\ne N$, \crcr}
\qquad
-L_k|_{S_N}=\cases{
\gamma_{N-1}+N\varepsilon_N & $k=N$ \crcr
k\varepsilon_N & $k< N$. \crcr}
}

It is then easy to see that $-L_k|_{S_j}$ is a nef divisor which maps
$S_j$ to a point for $j<k$, contracts $S_j$ along its ruling (mapping
to a curve) for $j>k$, is ample on the quadric or rational cone
$\overline{S}_k$ when $j=k<N$, and corresponds to the divisor
$\gamma_{N-1}+N\varepsilon_N=-K_{S_N}+(N-1)\varepsilon_N$ on the
surface $\overline{S}_N$ when $j=k=N$.

The key question now is: is the divisor $-K_{S_N}+(N-1)\varepsilon_N$
ample on the surface $\overline{S}_N$? We have not succeeded in
settling this question in general; however, we can construct an
example of such a surface on which the divisor is ample, as follows.
Start with the Hirzebruch surface $\IF_N$, and let $\sigma_\infty$ be
the section of self-intersection $-N$, and $f$ be the class of the
fiber.  The divisor class $2\sigma_\infty+(2N+2)f$ contains a smooth
curve $C$ of genus $N+1$ which meets $\sigma_\infty$ in two points,
and meets each fiber $f$ in two points.  Let $\pi:S\to \IF_N$ be the
double cover of $\IF_N$ branched along $C$, let
$\varepsilon=\pi^{-1}(f)$ and let $\gamma=\pi^{-1}(\sigma_\infty)$.
The surface $S$ is ruled by the rational curves $\varepsilon$, and
$\gamma$ is a $2$-section for this ruling which has self-intersection
$-2N$.  Moreover, the curves $\varepsilon$ are generically irreducible
but become reducible precisely when the corresponding fiber $f$ is
tangent to $C$, i.e., precisely where the two-to-one map
$C\to\sigma_\infty$ has branch points.  Since $C$ has genus $N+1$,
there are precisely $2N+4$ such branch points, and so precisely $2N+4$
line pairs as singular fibers of the ruling on $S$.  Thus, $S$
satisfies all of the properties needed for it to be used as $S_N$ in
our description of the configuration leading to $Sp(N)$ gauge theory
with $n_F=2N+4$.

It remains to verify that the divisor $-K_{S}+(N-1)\varepsilon$ is
ample on the surface $\overline{S}$ obtained {}from $S$ by contracting
$\gamma$.  By construction,
\eqn\kcalc{
-K_S+(N-1)\varepsilon=\pi^*(-K_{\IF_N}-\half C) + (N-1)\pi^*(f)
=\pi^*(\sigma_\infty+Nf).  } The divisor class $\sigma_\infty+Nf$ on
$\IF_N$ contracts $\sigma_\infty$ to a point, and embeds the resulting
surface $\overline{\IF}_N$ into $\IP^N$; it follows that its pullback
to $\overline{S}$ must be ample on that surface.

Since there exist examples of the surfaces $\overline{S}_N$ with the
required property, a strong coupling limit of the gauge theory will
exist.  This strongly suggests that any configuration of surfaces on a
Calabi--Yau manifold giving rise to that gauge theory should have a
strong coupling limit, and hence should have the property that
$-K_{S_N}+(N-1)\varepsilon_N$ is in fact ample on $\overline{S}_N$.

\subsec{$SU(N)$}

As our next example, we consider $SU(N)$ with $g=0$ adjoints and
$n_F=2N-2$ fundamentals, with $c_{classical}=0$.  Although this is not
quite the maximum number of fundamentals which we expect to be
possible based on our convexity analysis, analyzing the ampleness
condition is quite complicated in the case of $n_F=2N$---even more
complicated than the $Sp(N)$ case with $n_A=0$ which we discussed in
the previous subsection.  However, it turns out that concrete examples
with $n_F=2N-2$ can be given using toric geometry, so we shall present
those instead.

The toric data needed for these examples is illustrated in figure 5,
which shows a slice of a cone which could be used to describe an
affine toric variety.\foot{For a review of toric geometry, see \AGM.}
The solid circles represent the divisors $S_1$, \dots, $S_{N-1}$ which
should be contracted, the open circles represent other toric divisors
in a toric ambient space.  To perform the contraction of all of these
divisors to a point, one simply omits the solid circles and all
associated edges.

The configuration of the divisors is determined by the connections
between the dots, and there will be $N-1$ different triangulations
which must be considered, corresponding to the possibility of flopping
among different resolutions. (Both relevant triangulations in the
$SU(3)$ case are depicted in figure 5.)  An easy exercise in toric
geometry verifies that these configurations reproduce the data
contained in figure 2 and the surround description, albeit in a
slightly degenerate form in which the blowups {}from a minimal ruled
surface have all been concentrated at two points (which are each blown
up several times).  Also {}from this data, one checks easily that
$c'=1-N$, and hence that $c_{classical}=c'+{n_F\over2}=0$.

\iffigs
\midinsert
\centerline{\epsfxsize=1.75in\epsfbox{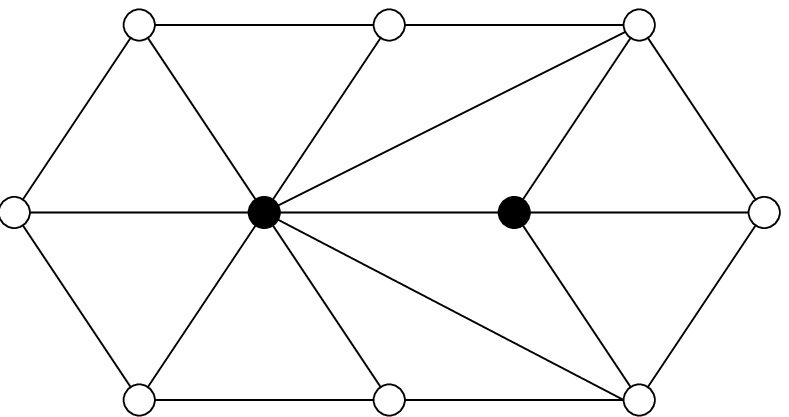}\qquad\qquad
\epsfxsize=1.75in\epsfbox{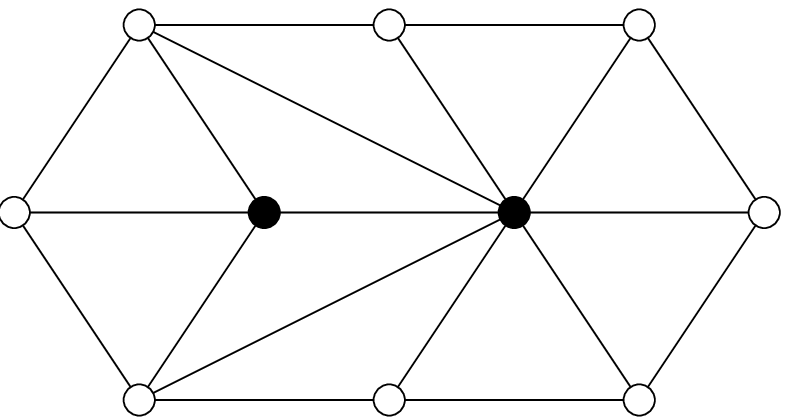}}
\centerline{Figure 5. Two triangulations of toric data for $SU(3)$ with
$n_F=4$.}
\endinsert
\fi

\subsec{$Spin(2N+1)$}

In our next example, we consider $Spin(2N+1)$ with $g=0$ adjoints and
$n_V=2N-3$ vectors.  The geometry is represented by Figure 3, with
$S_1$, \dots, $S_{N-1}$ being minimal rational ruled surfaces meeting
along rational curves $\gamma_j$ ($j\le N-2$), and $S_N$ being a
minimal ruled surface over a curve $\gamma_{N-1}$ of genus
$g'=2N-3$. {}From \aabsoodd, we see that
\eqn\sisoodd{
(\gamma_{j-1}^2)_{S_j} =S_{j-1}^2S_{j}
=\widetilde{S}_{j-1}^2\widetilde{S}_{j} =-2j,
}
for $1<j< N$, and
\eqn\sisooddbis{
(\gamma_{N-1}^2)_{S_N} =S_{N-1}^2S_{N}
=2\widetilde{S}_{N-1}^2\widetilde{S}_{N} =-4N.  } We can thus contract
$\gamma_{j-1}$ on $S_j$ to obtain a rational cone $\overline{S}_j$ for
$1<j<N$, and contract $\gamma_{N-1}$ on $S_N$ to obtain a cone
$\overline{S}_N$ over a curve of genus $2N-3$.  As before, $S_1$ can
be either $\IF_0$ or $\IF_2$; we let $\overline{S}_1$ denote either
$\IF_0$ or the contraction of the negative section on $\IF_2$.

The divisor $L_k:=\sum_{j=1}^{k-1}jS_j+\sum_{j=k}^{N-1}kS_j+\half
kS_N$ lies on that edge of the cone $-\cK(X/\overline{X})$ for which
$a_1=\dots=a_k>a_{k+1}=\dots=a_N=0$.  We define
$\gamma_0=\gamma_1-2\varepsilon_1$ as a class on $S_1$, and note that
$S_j\cong\IF_{2j}$ for
$1<j<N$, and hence that $\gamma_{j-1} +2j\varepsilon_j \sim \gamma_j$ on
$S_j$.  The normal bundles are $S_j|_{S_j}=K_{S_j}=
-\gamma_{j-1}-\gamma_j-2\varepsilon_j$ for $j<N$ and
$S_N|_{S_N}=K_{S_N}=-\gamma_{N-1}+4\varepsilon_N$, and the restriction
$-L_k|_{S_j}$ can be computed as follows:
\eqn\lksjsoodd{
-L_k|_{S_j}=\cases{0 & $k>j$\crcr
\gamma_{j} & $k=j<N$ \crcr
2k\varepsilon_j & $k<j\ne N$, \crcr}
\qquad
-L_k|_{S_N}=\cases{
\gamma_{N-1}+4N\varepsilon_N & $k=N$ \crcr
4k\varepsilon_N & $k< N$. \crcr}
}

It is then easy to see that $-L_k|_{S_j}$ is a nef divisor which maps
$S_j$ to a point for $j<k$, contracts $S_j$ along its ruling (mapping
to a curve) for $j>k$, is ample on the quadric or rational cone
$\overline{S}_k$ when $j=k<N$, and is ample on the irrational cone
$\overline{S}_N$ when $j=k=N$.

\subsec{$Spin(2N)$}

As our final example, we consider $Spin(2N)$ with $g=0$ adjoints and
$n_V=2N-4$ vectors.  The geometry is represented by Figure 4, with
$S_1$, \dots, $S_{N-1}$ being minimal rational ruled surfaces meeting
along rational curves $\gamma_j$ ($j\le N-2$), and $S_N$ being a
non-minimal rational ruled surface with double points, which meets
both $S_{N-1}$ and $S_{N-2}$ along rational curves.  More precisely,
$S_N$ is obtained {}from a minimal ruled surface by performing $n_V$
directed blowups of weight $1$, yielding exceptional divisors
$\sigma_N^{(\alpha)}$ passing through double points, and the proper
transforms $\delta^{(\alpha)}$ of the fibers passing through the points
that were blown up.

{}From \aabsoeven, we see that
\eqn\sisoeven{
(\gamma_{j-1}^2)_{S_j} =S_{j-1}^2S_{j}
=\widetilde{S}_{j-1}^2\widetilde{S}_{j} =-2j, } for $j\le N-1$. We can
thus contract $\gamma_{j-1}$ on $S_j$ to obtain a rational cone
$\overline{S}_j$ for $j\le N-1$.  As in some previous examples, $S_1$
can be either $\IF_0$ or $\IF_2$; we let $\overline{S}_1$ denote
either $\IF_0$ or the contraction of the negative section on $\IF_2$.

{}From \curvesoevenbis\ and \propE\ we see that
\eqn\sisoevenbis{
(\gamma_{N-2}^2)_{S_N} =S_{N-2}^2S_{N}=-2-n_V=2-2N.  } We can thus
contract $\gamma_{N-1}$ and all $\sigma_N^{(\alpha)}$'s (which are
disjoint {}from it) on $S_N$ to produce a rational cone
$\overline{S}_N$.  Alternatively, since $\gamma_{N-1}$ and the
$\delta^{(\alpha)}$'s have an intersection matrix
\eqn\intmatrix{
\pmatrix{-2-n_V & 1 & \cdots & 1 \crcr
1 & -2 &&\crcr
\vdots & &  \ddots &\crcr
1&&&-2}
}
which is negative definite, we can contract all of them to produce a
surface $\overline{\overline{S}}$, which is a sort of ``cone with double
points.''

The divisor $L_k:=\sum_{j=1}^{k-1}jS_j+\sum_{j=k}^{N-1}kS_j+\half
k(S_N-S_{N-1})$ lies on that edge of the cone $-\cK(X/\overline{X})$
for which $a_1=\dots=a_k>a_{k+1}=\dots=a_N=0$.  We define
$\gamma_0=\gamma_1-2\varepsilon_1$ as a class on $S_1$, and note that
$S_j\cong\IF_{2j}$ for
$1<j<N$, and hence that $\gamma_{j-1} +2j\varepsilon_j \sim \gamma_j$ on
$S_j$; also, $\gamma_{N-1}+(2N-4)\varepsilon_{N-2}
\sim \half\gamma_{N-2}+\half\gamma_{N-1}$ on $S_{N-2}$.  The normal bundles
are $S_j|_{S_j}=K_{S_j}=
-\gamma_{j-1}-\gamma_j-2\varepsilon_j$ for $j<N$ and
$S_N|_{S_N}=K_{S_N}=-\gamma_{N-1}-\delta+2\varepsilon_N$, and the restriction
$-L_k|_{S_j}$ can be computed as follows:
\eqn\lksjsoeven{
-L_k|_{S_j}=\cases{0 & $k>j$\crcr
\gamma_{j} & $k=j<N$ \crcr
2k\varepsilon_j & $k<j\ne N$, \crcr}
\qquad
-L_k|_{S_N}=\cases{
2\gamma_{N-1}+\delta+2N\varepsilon_N & $k=N$ \crcr
\gamma_{N-1}+(2N-2)\varepsilon_N & $k=N-1$ \crcr
2k\varepsilon_N & $k< N-1$, \crcr}
}
where $\delta=\sum\delta^{(\alpha)}$.

It follows that $-L_k|_{S_j}$ is a nef divisor which maps $S_j$ to a
point for $j<k$, contracts $S_j$ along its ruling (mapping to a curve)
for $j>k$ (except $(j,k)=(N,N-1)$), is ample on the quadric or
rational cone $\overline{S}_k$ when $j=k<N$ or $(j,k)=(N,N-1)$, and is
ample on the ``cone with double points'' $\overline{\overline{S}}_N$
when $j=k=N$.

Note that $-L_{N-1}$ might have been expected to leave only $S_{N-1}$
uncontracted, it in fact leaves both of the divisors $S_N$ and
$S_{N-1}$ uncontracted, mapping them to cones which meet along a common
fiber.  The reader may wonder if there is some divisor which contracts all
of the $S_j$'s other than $S_{N-1}$. Indeed there is, but the flop along
the $\sigma_N^{(\alpha)}$'s must be performed before that divisor belongs
to the relative K\"ahler cone.

\bigskip
\centerline{{\bf Acknowledgments}}

We would like to thank Paul Aspinwall, Tom Banks, Ron Donagi,
Antonella Grassi, Jeff Harvey, Sheldon Katz, Greg Moore, Miles Reid,
Steve Shenker, and Edward Witten for discussions.  D.R.M. would like
to acknowledge the hospitality of the Research Institute for
Mathematical Sciences, Kyoto University, where part of this work was
done.  The work of K.I. is supported by NSF PHY-9513835, the W.M. Keck
Foundation, an Alfred Sloan Foundation Fellowship, and the generosity
of Martin and Helen Chooljian.  The work of D.R.M. is supported in
part by the Harmon Duncombe Foundation and by NSF grants DMS-9401447
and DMS-9627351.  The work of N.S. is supported in part by DOE grant
\#DE-FG02-96ER40559.

\listrefs
\end